\documentclass[12pt,a4paper]{article}
\usepackage{type1cm,amsmath,graphicx,indentfirst}
\usepackage{amssymb}
\numberwithin{equation}{section}

\DeclareMathOperator{\str}{str}

\newcommand{\del}{\partial}
\newcommand{\delbar}{\bar{\partial}}

\begin{document}
\begin{titlepage}

 \renewcommand{\thefootnote}{\fnsymbol{footnote}}
\begin{flushright}
 \begin{tabular}{l}
 \end{tabular}
\end{flushright}

 \vfill
 \begin{center}


\noindent{\large \textbf{Three point functions in higher spin AdS$_3$ supergravity
}}\\
\vspace{1.5cm}

\noindent{ Thomas Creutzig,$^{a,b}$\footnote{E-mail: tcreutzig@mathematik.tu-darmstadt.de} Yasuaki Hikida$^c$\footnote{E-mail:
hikida@phys-h.keio.ac.jp} and Peter B. R\o nne$^{d}$\footnote{E-mail: peter.roenne@uni-koeln.de}}
\bigskip

 \vskip .6 truecm
\centerline{\it $^a$Fachbereich Mathematik,
Technische Universit\"{a}t Darmstadt,}
\centerline{\it Schlo\ss gartenstr. 7
64289 Darmstadt, Germany}
\medskip
\centerline{\it $^b$Hausdorff Research Institute for Mathematics,}
\centerline{\it{Poppelsdorfer Allee 45,
53115 Bonn,
Germany }}
\medskip
\centerline{\it $^c$Department of Physics, and Research and Education
Center for Natural Sciences,}
\centerline{\it  Keio University, Hiyoshi, Yokohama 223-8521, Japan}
\medskip
\centerline{\it $^d$Institut f\"{u}r Theoretische Physik, Universit\"{a}t zu K\"{o}ln,
} \centerline{\it
Z\"{u}lpicher Stra{\ss}e 77, 50937 Cologne, Germany}
 \vskip .4 truecm

 \end{center}

 \vfill
\vskip 0.5 truecm

\begin{abstract}

In a previous work we have proposed that the Prokushkin-Vasiliev higher spin ${\cal N}=2$ supergravity on AdS$_3$
is dual to a large $N$ limit of the ${\cal N}=(2,2)$ $\mathbb{C}$P$^N$ Kazama-Suzuki model.
There is now strong evidence supporting this proposal based on symmetry and spectrum comparison.
In this paper we will give further evidence for the duality by studying correlation functions.
We compute boundary three point functions with two fermionic operators and one higher
spin bosonic current in terms of the bulk supergravity theory.
Then we compare with the results in the dual CFT, where the supersymmetry of the theory
turns out to be very helpful.
In particular we use it to confirm results conjectured in the bosonic case.
Moreover, correlators with a fermionic current can be obtained via supersymmetry.

\end{abstract}
\vfill
\vskip 0.5 truecm

\setcounter{footnote}{0}
\renewcommand{\thefootnote}{\arabic{footnote}}
\end{titlepage}

\newpage

\tableofcontents

\section{Introduction}

In this paper, we study the behavior of massive fermions in the higher spin
${\cal N}=2$ supergravity on AdS$_3$ found by Prokushkin and Vasiliev \cite{PV1}.
{}From the behavior of these bulk fermions we compute boundary three point functions
with two fermionic operators and one higher spin bosonic current.
Higher spin gravity theories on AdS spaces have attracted a lot of attention, most importantly
for their application to the AdS/CFT correspondence.
In \cite{KP} (see \cite{Sezgin:2002rt} for a prior work) it was proposed that the Vasiliev higher spin gravity on AdS$_4$ \cite{Vasiliev:2003ev} is dual to the O$(N)$ vector model in three dimensions. 
There are many works related to this proposal and in particular some boundary
correlation functions were reproduced in terms of the dual gravity theory
\cite{Giombi:2009wh,Giombi:2010vg,Giombi:2011ya}.

For AdS$_3$ it was proposed by Gaberdiel and Gopakumar \cite{GG} (see \cite{Gaberdiel:2012uj} for a review)
that a truncated version of the higher spin
gravity by Prokushkin and Vasiliev is dual to a large $N$ limit of ${\cal W}_N$ minimal models
\begin{align} \label{minimal}
 \frac{\widehat{\text{su}}(N)_k \oplus \widehat{\text{su}}(N)_1 }{\widehat{\text{su}}(N)_{k+1} }
\end{align}
with the 't Hooft parameter
\begin{align} \label{thooft}
 \lambda = \frac{N}{N+k}
\end{align}
kept finite.
The proposal for the case related to the ${\cal WD}_N$ minimal model was presented in
\cite{Ahn:2011pv,Gaberdiel:2011nt}, and in \cite{CHR} we extended the conjecture to the full untruncated $\mathcal{N}=2$ supersymmetric case.%
\footnote{The $\mathcal{N}=1$ supersymmetric version of the duality is proposed in \cite{CHR3}.}
In this note we would like to give more evidence
supporting the proposal in \cite{CHR} by studying correlation functions.

There is already strong evidence in support of the proposal in \cite{GG}.
First of all, the asymptotic symmetry of the higher spin gravity has been
identified as a large $N$ limit of the ${\cal W}_N$ symmetry in \cite{HR,CFPT,Gaberdiel:2011wb,CFP,Gaberdiel:2012ku}.
This fact leads to the conjecture of \cite{GG} that the dual CFT is the 't Hooft limit of
${\cal W}_N$ minimal model \eqref{minimal}.
More impressively, the one-loop partition function of the gravity theory
was reproduced by the large $N$ limit of the dual CFT in \cite{Gaberdiel:2011zw}.
This means the duality was shown to be true in the free limit of the gravity theory.
In order to check the duality beyond the limit, we have to include interactions on the gravity side.
In fact, some boundary correlation functions were already investigated
in \cite{CY,Papadodimas:2011pf,Ahn:2011by,AKP,Chang:2011vk,Ahn:2012gw},
and up to now the results are consistent with the proposed duality.

For the full untruncated case in \cite{CHR}, the duality relates the ${\cal N}=2$ higher spin supergravity on AdS$_3$
found by Prokushkin and Vasiliev \cite{PV1} to the ${\cal N}=(2,2)$ $\mathbb{C}$P$^N$ Kazama-Suzuki model \cite{Kazama:1988qp,Kazama:1988uz}
\begin{align} \label{sminimal}
 \frac{\widehat{\text{su}}(N+1)_k \oplus \widehat{\text{so}}(2N)_1 }
{\widehat{\text{su}}(N)_{k+1}  \oplus \widehat{\text{u}}(1)_{N(N+1)(k+N+1)}}
\end{align}
in the large $N$ limit with the 't Hooft parameter \eqref{thooft} kept finite.
Also in this case there is strong evidence to support the conjecture.
As in the bosonic case, the asymptotic symmetry
of the supergravity is found to be a large $N$ limit of the ${\cal N}=(2,2)$ ${\cal W}_N$
algebra \cite{CHR,Henneaux:2012ny,Hanaki:2012yf,Candu:2012tr}.
Given this fact, the most plausible candidate is the ${\cal N}=(2,2)$ ${\cal W}_N$
minimal model which can be described by the $\mathbb{C}$P$^N$ Kazama-Suzuki model
\cite{Ito}. Moreover, the one-loop partition function of the supergravity
is reproduced by the 't Hooft limit of the Kazama-Suzuki model
\cite{CG}. We can thus conclude that the spectra of the dual theories agree.
So the next task should be to examine  boundary correlation functions.
For AdS$_4$/CFT$_3$ as in \cite{KP}, it was argued in \cite{MZ1,MZ2}
that the correlation functions are quite restricted due to the higher spin symmetry.
Even with this fact,
it was also pointed out in these papers that for examples of AdS$_3$/CFT$_2$ the
higher spin symmetries are not so restrictive, and extra studies are needed.
See \cite{Fredenhagen:2012rb,Ahn:2012fz,Ahn:2012vs,Tan:2012xi,Datta:2012km,Fredenhagen:2012bw} for recent developments on the ${\cal N}=2$ minimal model holography.

The supergravity theory of \cite{PV1} consists of massless higher spin gauge fields and massive
matter fields. There are two sets of bosonic gauge fields with respectively spins $s=1,2,\ldots$ and $s=2,3,\ldots$, and two sets of fermionic gauge fields both with spins $s=3/2,5/2,\ldots$.
The dual currents we denote by $J^{(s)\pm}$.
There are also four massive complex scalar fields and four massive Dirac spinor fields with spin $1/2$.
The dual operators ${\cal O}^{(h,\bar h)}$ may be labeled by their conformal weights
$(h , \bar h)$. For the bosonic operators the conformal weights satisfy $h = \bar h$,
 and for the fermionic operators they are $h = \bar h \pm 1/2 $.
In this paper we compute boundary three-point functions with two fermionic operators
${\cal O}_F^{(h,\bar h)}$ and one bosonic higher spin current $J^{(s)\pm}$ i.e.
\begin{align}
 \langle {\cal O}_F^{(h , \bar h)} (z_1) \widetilde  {\cal O}_F^{(h , \bar h)} (z_2) J^{(s)\pm} (z_3) \rangle
 \label{3pti}
\end{align}
where $s$ is integer. In the bosonic case where
$h = \bar h $ the
three-point functions have been computed in \cite{CY} with a restricted parameter
$\lambda = 1/2$, and later in \cite{AKP} with arbitrary $\lambda$ using a simple
method. Here we apply the method of \cite{AKP} for the computation. We find that the structure constants for
the correlators of the fermionic operators are the same as for the bosonic correlators.

We then examine the results obtained in the bulk from the viewpoint of the dual CFT, and we explain the bulk results
via supersymmetry. On the bulk side there is a simple relation between the two bosonic higher spin currents $J^{(s)\pm}$
when acting on the matter states, namely $J^{(s)-}=\pm J^{(s)+}$. Assuming this in the CFT, we obtain, via supersymmetry,
a simple recursion relation between correlators with currents of spin $s$ and $s+1$.
{}From this relation we can reproduce exactly the conjectured results of \cite{AKP}. Further, we explicitly construct the higher spin currents up to spin 2 in the super coset theory, and show that
the spin two currents indeed have the simple relation when acting on the dual matter states. Finally, we show that the found currents
are the generating currents for the whole super $\mathcal{W}[\lambda]$ algebra.

This paper is organized as follows.
In the next section we review the ${\cal N} =2$ higher spin supergravity constructed in \cite{PV1}.
We are then set for section \ref{fermions} where we study the behavior of the massive fermions in the supergravity
with AdS background. In section \ref{3ptfn} we compute three point functions of the form \eqref{3pti} with two fermionic operators and one bosonic higher spin current from the viewpoint of the bulk theory. To prepare for the CFT analysis we study how the supersymmetry transformations and states of the bulk theory
map to currents and operators of the boundary theory in section \ref{dictionary}.
In section \ref{CFT} we explain the results obtained from the bulk supergravity via the supersymmetry structure of the dual CFT. Further, we obtain the recursion relation for the correlators, and provide strong support for the validity of it.
Finally, we conclude in section \ref{conclusion}.
In appendix \ref{hsas}, structure constants of the higher spin algebras hs[$\lambda$] and shs[$\lambda$] are reviewed.
Some explicit computations involving the star product have been done in appendix \ref{structure}, and also the (anti-)automorphisms and the supertrace of the algebra can be found there.
In appendix \ref{CFTW}, operator products in a CFT with ${\cal N}=2$ super ${\cal W}$ symmetry algebra are summarized.

\subsection*{Note added}

After putting this draft in its final form, we were informed that boundary three-point functions in the ${\cal N}=2$
higher spin holography are also analyzed in \cite{MZ}. In their paper the three-point functions
considered from the bulk side are those with two bosonic operators and one bosonic higher spin current,
however, calculated using an alternate basis. In our paper we additionally deal with those three-point functions having two fermionic operator insertions.

\section{Higher spin AdS$_3$  supergravity}
\label{vasiliev}

In \cite{PV1} a higher spin ${\cal N}=2$ supergravity theory in three
dimensions has been developed where massive scalars and fermions are
coupled with higher spin gauge fields.
Field equations are given in the paper, but the action of the theory is not known yet.
We are interested in a vacuum solution with AdS$_3$ space and small deformations thereof, and
in this case we can use the $\text{shs}[\lambda] \otimes \text{shs}[\lambda]$ Chern-Simons
gauge theories coupled with massive matter. We only explain the results here briefly since
the arguments are essentially the same as
in \cite{CY,AKP}, but now without the truncation to bosonic subsector.%
\footnote{See, for instance, appendix A of \cite{AKP} for a good review.}

\subsection{Supergravity by Prokushkin and Vasiliev}

The supergravity theory consists of the generating functions  $(W_\mu,B,S_\alpha)$.
The space-time one-form $W = W_\mu d x^\mu$ and the zero form $B$
describes the massless higher spin gauge fields and the massive matter fields,
respectively. The fields $S_\alpha$ are auxiliary, and they generate constraints of the other fields.
Here and in the following $\alpha=1,2$ is the spinor index and
it is raised and lowered by the antisymmetric tensors
$\epsilon_{12} = \epsilon^{12} = 1$.
The generating functions depend on the generators
$(z_\alpha,y_\alpha,\psi_{1,2},k,\rho)$ as well as the space-time coordinates $x^\mu$. These generators fulfill the following algebraic relations
\begin{align}
 k^2 = \rho^2 = 1 ~, \qquad \{ k , \rho \} = \{ k , y_\alpha \} =
\{ k , z_\alpha \} = 0 ~, \qquad \{\psi_i,\psi_j\}=2\delta_{ij}
\end{align}
with all the remaining commutators being zero. 
The fields of the theory are obtained by
expanding the generating functions as
\begin{align}
 A(z,y,\psi_{1,2},k,\rho | x) = \sum_{B,C,D,E=0}^1 \sum_{m,n=0}^\infty
  A^{BCDE}_{\alpha_1 \ldots \alpha_m  \beta_1 \ldots \beta_n }(x) k^B \rho^C \psi_1^D \psi_2^E
  z_{\alpha_1} \ldots z_{\alpha_m} y_{\beta_1} \ldots y_{\beta_n} ~.
\end{align}

The product of generating functions in terms of the twistor variables $z_\alpha ,y_\alpha$ is
defined by the star product
\begin{align}\label{moyalstar}
 (f * g) (z,y) = \frac{1}{(2\pi)^2}
 \int d^2 u d^2 v
 e^{i u_\alpha v^\alpha }
 f(z + u , y + u) g (z - v , y + v) ~.
\end{align}
With this product law, the field equations are \cite{PV1}
\begin{align}\label{fieldeq}
 &d W = W * \wedge W ~, \qquad
 d B = W * B - B * W  ~, \\
 &d S_\alpha = W * S_\alpha - S_\alpha * W ~, \qquad
 S_\alpha * S^\alpha = - 2 i (1+B * K) ~, \qquad
 S_\alpha * B = B * S_\alpha ~,\nonumber
\end{align}
where
\begin{align}
 K = k  e^{i z_\alpha y^\alpha }
\end{align}
is called the Kleinian.
These  equations are invariant under the following higher spin gauge transformations
\begin{align}\label{eq:gaugetransoriginalfields}
 \delta W = d \varepsilon - W * \varepsilon + \varepsilon * W ~,  \qquad
 \delta B = \varepsilon * B - B * \varepsilon ~,  \qquad
 \delta S_\alpha = \varepsilon * S_\alpha - S_\alpha * \varepsilon ~,
\end{align}
where the gauge parameter $\varepsilon = \varepsilon (z,y; \psi_{1,2},k|x)$ is arbitrary, but $\rho$-independent.
Using the symmetry of the field equations under
$\rho \to - \rho, S_\alpha \to - S_\alpha$, we consider a truncated
system where $W_\mu,B$ are independent of $\rho$, and $S_\alpha$ is
linear in $\rho$.

We consider vacuum solutions of \eqref{fieldeq} denoted by $B_0, W_0, S_{0\alpha}$. We solve the equation of motion for $B$ by setting $B_0$ equal to a constant
\begin{align}
 B_0 = \nu ~.
 \label{vev}
\end{align}
The field equations then reduce to
\begin{align}
\label{fieldeq2}
 d W_0 = W_0 * \wedge W_0 ~, \qquad
 d S_{0\alpha} = W_0 * S_{0\alpha} -  S_{0\alpha}  * W_0 ~, \qquad
  S_{0\alpha} * S_0^\alpha = - 2 i (1 + \nu K ) ~.
\end{align}
A solution for $S_{0\alpha}$ is given by
\begin{align}
 S_{0\alpha} = \rho \tilde z_\alpha ~,
\end{align}
where
\begin{align}
 \tilde z_\alpha = z_\alpha + \nu w_\alpha k ~, \qquad
 w_\alpha = (z_\alpha + y_\alpha) \int_0^1 dt \, t e^{it z_\alpha y^\alpha} ~.
\end{align}
It is convenient also to define $\tilde y_\alpha$ as
\begin{align}
 \tilde y_\alpha = y_\alpha + \nu w_\alpha * K ~, \qquad
 [\tilde y_\alpha , \tilde y_\beta ]_* = 2 i \epsilon_{\alpha \beta} (1+\nu k) ~, \qquad
 [\rho \tilde z_\alpha , \tilde y_\beta]_* =0
\end{align}
with $[A,B]_* = A*B-B*A$.
Since $d S_{0\alpha}=0$, generic solutions for $W_0$ have to commute with $\rho \tilde z_\alpha$, i.e. they are given by functions
of the generators $k,\tilde y_\alpha$ and $\psi_{1,2}$, but are independent of $\tilde z_\alpha$. The only remaining field equation is then
the first equation of \eqref{fieldeq2}.

\subsection{Higher spin gauge fields}

As found in the previous subsection, the vacuum value of $W=W_0$ is parameterized
by $\psi_{1,2},k,\tilde y_\alpha$ and the space-time coordinates $x_\mu$.
It was shown in \cite{CY} that the part linear in $\psi_2$ is
auxiliary, so we can neglect $\psi_2$. Now that $\psi_1$ commutes with
all variables and $\psi_1^2=1$, we may define projection operators
\begin{align}
 \Pi_\pm = \frac{1 \pm \psi_1}{2} ~.
\end{align}
Then we can rewrite the field equation for $W_{0}$ as
\begin{align}
 d A + A * \wedge A = 0 ~, \qquad d \bar A + \bar A * \wedge \bar A = 0
 \label{CSeq}
\end{align}
with
\begin{align}
 W_0 = - \Pi_+ A - \Pi_- \bar A ~.
\end{align}
Here $A$ and $\bar A$ are functions of $\tilde y_\alpha$ and $k$.
The above field equations for $A,\bar A$ are the same as the equations of motion for
Chern-Simons theory based on the algebra generated by $\tilde y_\alpha$ and $k$.

Before discussing the full algebra with $\tilde y_\alpha,k$, let us review the bosonic
truncation where we only allow an even number of $\tilde y_\alpha$ in the generators and we project onto one of the two possible eigenvalues $k= +1$ or $k=-1$ (which is allowed since $k$ is commuting with an even number of $\tilde y_\alpha$).
In this case, the algebra is called hs[$\lambda_\pm$] (see, e.g., \cite{Gaberdiel:2011wb}) where $\lambda$ depends on the choice of eigenvalue of $k$
\begin{align}
\lambda_\pm = \frac{1 \mp \nu }{2}\quad\textrm{for }k=\pm1 ~. \label{lambdak}
\end{align}
The generators of hs[$\lambda$] are given by $V_m^s$ with $s = 2, 3, \ldots$ and
$|m| = 0, 1, \ldots ,s-1$. The commutation relations are
\begin{align}
 [ V_m^s , V_n^t] = \sum_{u=2,4,\cdots}^{s+t - |s-t|-1} g_{u}^{st} (m,n;\lambda) V_{m+n}^{s+t-u}
\end{align}
with the structure constant given in \eqref{gust}.  In particular, $V_m^2$ with $m=0,\pm 1$ generate
the sl$(2)$ subalgebra.
In order to compute star products among the generators $V_m^s$, we utilize the lone star product
defined in \cite{PRS} as
\begin{align}
 V_m^s * V_n^t = \frac12 \sum_{u=1,2,\cdots}^{s+t-|s-t| -1}
 g^{st}_u (m,n;\lambda) V^{s+t-u}_{m+n} ~.
 \label{lsp}
\end{align}
Indeed, it was conjectured in \cite{AKP} that the
generators are expressed in terms of $\tilde y_\alpha$ as
\begin{align}
 V_m^s = \left( \frac{-i}{4}\right)^{s-1} S_m^s ~.
 \label{symy}
\end{align}
Here $S^s_m$ is the symmetrized product of generators $\tilde y_\alpha$ where the total number of
generators $\tilde y_\alpha$ is $2s - 2$ and $2 m = N_1 - N_2$ with
the number of $\tilde y_{1,2}$ given by $N_{1,2}$. The precise normalization is
\begin{align}\label{}
    S^s_m=\frac{1}{(2s-2)!}\sum_{\sigma\in S_{2s-2}}y_{\alpha_{\sigma(1)}}*\cdots *y_{\alpha_{\sigma(2s-s)}}\ ,
\end{align}
where $S_{2s-s}$ represents the $(2s-2)$-th symmetric group.
The previously defined (Moyal) star product \eqref{moyalstar} then maps to the lone star product as has been checked explicitly up to spin 4 in \cite{AKP}. Star products among the generators $S^s_m$ are then found directly via the lone star product \eqref{lsp}
without tedious computations to symmetrize the products.

We now turn to the full algebra  where we can have both even and odd numbers of generators $\tilde y_\alpha$ and $k$-dependence. This algebra
was analyzed in \cite{Bergshoeff:1990cz,Bergshoeff:1991dz}, see also appendix \ref{hsas}.
We choose to denote the algebra shs[$\lambda$] where $\lambda$ is related to the vacuum expectation value $\nu$ as
\begin{align}\label{nulambda}
    \nu=1-2\lambda ~.
\end{align}
Again, for uniqueness, we choose generators that are symmetric products of the generators $\tilde y_\alpha$ now possibly multiplied with $k$.
As above, we denote these symmetric products $S^s_m$ where the even case has $s\in\mathbb{N}$ and $m\in \mathbb{Z}$, and the odd case has $s\in\mathbb{N}+1/2$ and $m\in \mathbb{Z}+1/2$, and we always have $|m| \leq s - 1$. We can now write our generators in the notation \eqref{symy} as
\begin{align}
 V^{(s)+}_m = \left( \frac{-i}{4}\right)^{s-1} S^s_m ~, \qquad
 V^{(s)-}_m &=  \left( \frac{-i}{4}\right)^{s-1} k S^s_m ~, \quad(s = 1, 3/2, 2,5/2,3,\ldots)\ .
\label{bosong}
\end{align}
For the spin algebra we have to pay special attention to the spin 1 case since we do not want to keep an overall central element. We thus only keep $k+\nu$ which appears in the anti-commutator $\{k \tilde y_\alpha,\tilde y_\beta\}=2i\epsilon_{\alpha\beta}(k+\nu)$. Some (anti-)commutation relations can be found in appendix \ref{hsas}. In this case, $\{k+\nu ,V^{(2)+}_m ,  V^{(3/2)\pm}_{m}\}$
generate the osp$(2|2)$ subalgebra, or in other words, the ${\cal N}=2$ supersymmetry,
see (10.4) of \cite{PV1}.

We note that the bosonic subalgebra splits into two subalgebras using projection operators onto the two eigenvalue spaces of $k$
\begin{align}
 P_\pm = \frac{1 \pm k}{2} ~. \label{pk}
\end{align}
Now the generators $P_+ S^s_m$ and $P_- S^s_m$ for $s=2,3,\ldots$ form respectively the algebras hs[$\lambda$] and hs[$1-\lambda$] and are mutually commuting due to the projectors. They correspond to the two bosonic subalgebras in the analytic continuation of sl$(N+1|N)$, see \cite{FL}.

The lone star product in \eqref{lsp} can be extended to the case with half-integer spin, but the expression is useless since the structure constants have not been
obtained at least in a simple form. The first few terms are computed in appendix \ref{structure}.
In other cases we use the bosonic version of \eqref{lsp} and multiplication of $V_{\pm 1}^{3/2}$,
as we will see below. In fact, generic structure constants should be computable in the
same way.

\subsection{Perturbation with massive matter}
\label{perturbations}

Up to now we only examined vacuum solutions, but
here we would like to discuss the perturbation with massive matter.
For this purpose we expand the generating function
$B$ around the vacuum value as
\begin{align}
 B = \nu + {\cal C} ~.
\end{align}
Then from the field equations \eqref{fieldeq} we have equations involving ${\cal C}$
\begin{align}
 d {\cal C} - W_0 * {\cal C} + {\cal C} * W_0 = 0 ~, \qquad
 [ S_{0\alpha}, {\cal C}]_* = 0 ~. \label{fieldeq3}
\end{align}
As for $W_0$, the second equation leads to ${\cal C}$ being a function of $\tilde y_\alpha$
and not of $\tilde z_\alpha$. Thus the perturbation can be written out as
\begin{align}
 B = \nu + \psi_2 {\cal C} (x_\mu , \tilde y_\alpha , k ) ~.
\end{align}
Here we neglect the part independent of $\psi_2$ since it only includes
auxiliary fields, see \cite{PV1}. As before, we decompose the fields into two
parts as
\begin{align}
 {\cal C} = \Pi_+ C(x_\mu , \tilde y_\alpha , k) \psi_2 +
 \Pi_-  \tilde C(x_\mu , \tilde y_\alpha , k) \psi_2~.
 \label{tildeC}
\end{align}
Then the first equation \eqref{fieldeq3} reduces to two equations
\begin{align}
 d C  + A * C - C * \bar A = 0 ~, \qquad d \tilde C + \bar A * \tilde C
 - \tilde C * A = 0 ~.  \label{feforc}
\end{align}

Considering the dependence on the variable $k$, we can
decompose the fields further using the projection operators \eqref{pk}
\begin{align}
 C = P_+ C_+ (x_\mu , \tilde y_\alpha) + P_- C_- (x_\mu , \tilde y_\alpha) ~,
 \qquad
 \tilde C = P_+ \tilde C_+ (x_\mu , \tilde y_\alpha)
 + P_- \tilde C_- (x_\mu , \tilde y_\alpha) ~.
\end{align}
The fields $C_\pm , \tilde C_\pm$ are polynomials of symmetric products of
$\tilde y_\alpha$, so they may be expanded as
\begin{align}
  C_\pm = \sum_{s=1,\frac{3}{2},2,\frac{5}{2},\ldots} \sum_{|m| \leq s-1} C^s_{m,\pm} V^s_m ~, \qquad
  \tilde C_\pm = \sum_{s=1,\frac{3}{2},2,\frac{5}{2},\ldots} \sum_{|m| \leq s-1} \tilde C^s_{m,\pm} V^s_m ~.
  \label{defc}
\end{align}
The Grassmann parity of the coefficients is discussed in (5.6) of \cite{PV1}
and in our notation integer $s$ components are Grassmann even and
half integer $s$ components are Grassmann odd as expected.

As shown in \cite{PV1}, any dynamics are described by $C^1_{0,\pm}$ and
 $\tilde C^1_{0,\pm}$ for bosonic modes and $C^{3/2}_{a,\pm}$ and
 $\tilde C^{3/2}_{a,\pm}$ for fermionic modes, where $a = \pm 1/2$.
If we consider the AdS vacuum, then the field equations for $C^1_{0,\pm},\tilde C^1_{0,\pm}$ reduce to
the Klein-Gordon equations  with masses
\begin{align}
 M^2_\pm = -1 + \lambda_\pm ^2  ~,
 \label{bosonmass}
\end{align}
where $\lambda_\pm = \tfrac12 ( 1 \mp \nu )$ as in \eqref{lambdak}. Thus the parameter $\nu$ enters the mass formula.
For $C^{3/2}_{a,\pm}, \tilde C^{3/2}_{a,\pm}$ the field equations reduce to the Dirac equations  with masses
\begin{align}
 M^2_\pm = (\lambda_\pm - \tfrac12)^2 ~,
 \label{fermionmass}
\end{align}
see (3.22) and (3.23) of \cite{PV1}.
Following the analysis for the scalars in \cite{AKP}, we re-derive the
Dirac equation with mass in the next section.

\section{Massive fermions on the AdS background}
\label{fermions}

Among the vacuum solutions of the field equations for supergravity,
the vacuum corresponding to AdS space plays a particular role due to its
application to the AdS/CFT correspondence. In this section, we study the
behavior of massive fermions on the AdS background. In the next section,
we introduce small deformations of the AdS background by introducing
non-vanishing higher spin fields.

\subsection{Dirac equations for the massive fermions}

Let us examine the field equation for $C$ \eqref{feforc} on the Euclidean AdS
background. We use the coordinate system $(\rho,z,\bar z)$,
where $\rho$ represents the radial direction of the AdS space
and its boundary is at $\rho \to \infty$.
The boundary coordinates are give by $z,\bar z$.
In these coordinates the AdS background has the metric
\begin{align}
  ds^2 = d \rho^2 + e^{2 \rho} d z d \bar z\ ,
\end{align}
which in turn corresponds to the following configuration
(see, e.g., eq. (3.8) of \cite{AKP})
\begin{align}
 A = e^\rho V_1^2 dz + V_0^2 d \rho ~, \qquad
\bar A = e^\rho V_{-1}^2 d \bar z - V_0^2 d \rho ~.
\label{background}
\end{align}
Here we have used the following relation between the frame-like and the metric-like formulation
\begin{align}
 e = \tfrac{1}{2} (A - \bar A) ~, \qquad g_{\mu \nu} \propto \text{tr} (e_\mu e_\nu) ~.
\end{align}
Since the above configuration only involves bosonic components,
we can truncate the label $s$ in \eqref{defc} to $s \in \mathbb{Z}$
or $s \in \mathbb{Z} + 1/2$. The former case is analyzed in \cite{AKP}.
Below we focus on $C^s_{a,\pm}$, but $\tilde C^s_{a,\pm}$ can be analyzed
in the same way.

With the above background,
the field equation \eqref{feforc} expressed in terms of the modes $C_m^s$ becomes (using the results of appendix \ref{structure})%
\footnote{Here we have suppressed the subscript $\pm$ in $C_{m , \pm }^s$. The dependence
only appears through $\lambda_\pm$.}
\begin{align}
 & \partial_\rho C^s_m + 2 C^{s-1}_m + h^s_m C^s_m + g^{(s+1)2}_3 (m,0) C^{s+1}_m = 0 ~,
\label{fe1}\\
 & \partial C^s_m + e^{\rho} (C^{s-1}_{m-1} + \tfrac12 g^{2s}_2 (1,m-1) C^s_{m-1}
 + \tfrac12 g^{2 (s+1)}_3 (1,m-1) C^{s+1}_{m-1} ) = 0 ~ , \label{fe2}\\
 & \bar \partial C^s_m - e^\rho (C^{s-1}_{m+1}
 + \tfrac12 g^{s2}_2 ( m+1 , - 1) C^s_{m+1}
 + \tfrac12 g^{ (s+1)2}_3 (m+1 , -1) C^{s+1}_{m+1}) = 0 \label{fe3}~,
\end{align}
where
\begin{align}
 h^s_m = \frac12 (g^{s2}_2 (m,0) + g^{2s}_ 2(0,m)) =
\left \{
\begin{array}{ll}
 0 & \text{for~} s \in \mathbb{Z} ~, \\
  \frac{m (1 - 2 \lambda_\pm)}{4s (s-1)} & \text{for~} s \in \mathbb{Z} + \tfrac12 ~.
  \end{array}
  \right .
 \label{hsm}
\end{align}
For integer $s$ the field equation \eqref{feforc} reduces to (3.10)
of \cite{AKP}.
 For half integer $s$ the equations are quite different since $h^s_m\neq0$ and
the functions $g^{2s}_3 (n,m) $
are also different from those with integer $s$ as shown in appendix \ref{structure}.
By a change of basis, we
can see that these equations reproduce (3.21) of \cite{PV1}.

First let us consider the case with integer $s$. {}From the whole set of equations,
we obtain a closed set $(C_0^1 , C_0^2 , C_0^3 , C_1^2 )$ as
\begin{align}
& \partial_\rho C_0^1 + \tfrac{\lambda_\pm^2 - 1}{6} C_0^2  = 0 ~, \qquad
 \bar \partial C_0^1 + e^\rho \tfrac{\lambda_\pm^2 - 1}{6} C_1^2  = 0 ~, \\
& \partial C_1^2 + e^\rho C_0^1 + \tfrac12 e^\rho C_0^2 - e^\rho  \tfrac{\lambda_\pm^2 - 4}{30} C_0^3 = 0 ~, \qquad
 \partial_\rho C_0^2 + 2 C_0^1 +\tfrac{2 (\lambda_\pm^2 - 4)}{15} C_0^3  = 0 ~. \nonumber
\end{align}
Solving these equations, we obtain the Klein-Gordon equation for $C_0^1$
\begin{align}
 [\partial_\rho^2 + 2 \partial_\rho + 4 e^{-2 \rho} \partial \bar \partial - (\lambda_\pm^2 - 1)]
 C_0^1 = 0 ~,
\end{align}
which leads to the mass formula
\begin{align}
 M_\pm^2 = - 1 +\lambda^2_\pm = - 1 + (\tfrac{1 \mp \nu}{2})^2
\end{align}
as mentioned in \eqref{bosonmass}.

Setting $(s,m)=(3/2,\pm 1/2)$ in  equations  \eqref{fe1}, \eqref{fe2} and \eqref{fe3},
we obtain another closed set $(C^{3/2}_{\pm 1/2},C^{5/2}_{\pm 1/2})$
\begin{align}
& \partial C^{\frac{3}{2}}_{\frac12} + e^\rho
 \left(\tfrac12 (1-\tfrac{1-2\lambda_\pm}{3})  C^{\frac{3}{2}}_{- \frac12}
 - \tfrac{(\lambda_\pm - 2)(\lambda_\pm + 1)}{18} C^{\frac{5}{2}}_{- \frac12}\right) = 0  ~, \\
&\bar  \partial C^{\frac{3}{2}}_{-\frac12} - e^\rho
 \left(\tfrac12 (1+\tfrac{1-2\lambda_\pm}{3})  C^{\frac{3}{2}}_{ \frac12}
 - \tfrac{(\lambda_\pm - 2)(\lambda _\pm+ 1)}{18} C^{\frac{5}{2}}_{ \frac12} \right)= 0  ~, \\
&\partial_\rho  C^{\frac{3}{2}}_{\frac12} + \tfrac{1 - 2 \lambda_\pm}{6}C^{\frac{3}{2}}_{\frac12}
 +  \tfrac{(\lambda _\pm- 2)(\lambda_\pm + 1)}{9} C^{\frac{5}{2}}_{\frac12} = 0 ~ , \\
&\partial_\rho  C^{\frac{3}{2}}_{-\frac12} - \tfrac{1 - 2 \lambda_\pm}{6}C^{\frac{3}{2}}_{-\frac12}
 +  \tfrac{(\lambda_\pm - 2)(\lambda_\pm + 1)}{9} C^{\frac{5}{2}}_{-\frac12} = 0 ~.
\end{align}
Eliminating $C^{5/2}_{\pm 1/2}$ we have
\begin{align}
&  (\partial_\rho +1 )  C^{\frac{3}{2}}_{-\frac12} + 2 e^{-\rho}
\partial  C^{\frac{3}{2}}_{\frac12} + ( \lambda_\pm - \tfrac12)
 C^{\frac{3}{2}}_{-\frac12} = 0 ~ , \nonumber \\
& - ( \partial_\rho +1 )  C^{\frac{3}{2}}_{\frac12} + 2 e^{-\rho}
\bar \partial  C^{\frac{3}{2}}_{-\frac12} + ( \lambda_\pm - \tfrac12)
 C^{\frac{3}{2}}_{\frac12} = 0 ~ .
 \label{dirac0}
\end{align}
These are nothing but the Dirac equations with mass
\begin{align}
M_\pm = \tfrac12 - \lambda_\pm
\end{align}
as in \eqref{fermionmass}.
We can repeat the same analysis for $\tilde C^{3/2}_{\pm 1/2}$, or simply use the anti-automorphism \eqref{eq:antiautoonCA},
and obtain the Dirac equations, but now the mass is
\begin{align}
 M_\pm = \lambda_\pm - \tfrac12
\end{align}
i.e. with $\lambda_\pm\mapsto\lambda_\mp$ or, equivalently, with the opposite sign.

\subsection{Solutions to the Dirac equation}

{}From the solutions to the Dirac equation, we can compute boundary correlation functions of the dual operators ${\cal O}^{[\delta]}_{F\pm}$. As in the bosonic case there are two types of boundary behaviour which we denote in the superscript by $\delta=\pm$. The subscript $\pm$ is again just referring to the $k$-projection and we will suppress it in the following.
The simplest case is the two point function of fermionic operators
\begin{align}
 \langle {\cal O}^{[\delta]}_F (z_1)   \widetilde{\cal O}^{[\delta]}_F (z_2) \rangle  ~.
\end{align}
We have here used that tilded and untilded fields couple, see eq. \eqref{eq:lagrangianstructure}. Note that this is basically due to the U$(1)$ symmetry of the $\mathcal{N}=2$ superalgebra. Using a more familiar notation $C^{3/2}_{\pm 1/2} = \psi_{\pm}$, the Dirac equation
\eqref{dirac0} becomes
\begin{align}
 (\partial_\rho + 1 + M) \psi_+ - 2 e^{- \rho} \bar \partial \psi_- = 0 ~,
 \qquad
 (\partial_\rho + 1 - M) \psi_- + 2 e^{- \rho}  \partial \psi_+ = 0 ~. \label{diracp}
\end{align}
A direct computation shows that
\begin{align}
   &\psi_+  (\rho , z ) = - \frac{M+\frac12}{\pi} \int d^2 z ' e^{ \frac12 \rho}
 \left ( \frac{e^{-\rho} }{e^{-2 \rho} + |z - z '|^2}\right)^{M+\frac32}
  (z-z') \eta_- (z ') ~, \label{psiplus}\\
 &\psi_-  (\rho , z ) = \frac{M+\frac12}{\pi} \int d^2 z ' e^{- \frac12 \rho}
 \left ( \frac{e^{-\rho} }{e^{-2 \rho} + |z - z '|^2}\right)^{M+\frac32}
  \eta_- (z ') \label{psiminus}
\end{align}
satisfy the Dirac equation, where $\eta_- (z')$ is a fermionic variable.
Around $\rho \sim \infty$, the solutions behave as
\begin{align}
 \psi_+ (\rho , z ) \sim 0 ~, \qquad \psi_- (\rho , z ) \sim
\eta_- (z) e^{\rho (M-1)} ~. \label{boundaryc}
\end{align}

Using the usual recipe of the AdS/CFT correspondence, we assign the boundary conditions for the fermions as
\begin{align}
  \psi_+ (\rho , z ) \sim 0 ~, \qquad \psi_- (\rho , z ) \sim
\varepsilon_- \delta^{(2)} (z - z_2 )e^{\rho (M-1)} ~,
\end{align}
where $\varepsilon_-$ is a constant parameter now. Then the two point function can be read off from the
solutions as
\begin{align}
 {\cal O} (z_1) = \varepsilon_- \langle {\cal O}_F(z_1)   \widetilde{\cal O}_F (z_2) \rangle  + \cdots
\end{align}
with
\begin{align}
 \psi_+ (\rho , z ) \sim \frac{{\cal O}(z)}{B_\psi} e^{ - \rho (M+1)}~,
 \qquad \psi_- (\rho , z ) \sim 0
\end{align}
around $\rho \sim \infty$ and $z \neq z_2$. Here $B_\psi$ represents the coupling between the bulk fermion and the boundary operator.
With this procedure, we can obtain the boundary two point function as
\begin{align}
 \langle {\cal O}^{[-]}_F(z_1)   \widetilde{\cal O}^{[-]}_F (z_2) \rangle
 = - \frac{B^{[-]}_\psi (M + \frac12)}{\pi} \frac{1}{z_{12}^{2h} \bar z_{12}^{2 \bar h}} ~.
 \label{2pt0}
\end{align}
{}Where the conformal weights of the dual fermionic operator are
$(h,\bar h)\equiv (h^{[-]} , \bar h^{[-]}) =(\frac{M+1/2}{2}, \frac{M+3/2}{2})$.
Inserting $M_\pm = \frac12 - \lambda_\pm$, it becomes
$(h^{[-]} , \bar h^{[-]}) = (\frac{1 - \lambda_\pm}{2} , \frac{2 - \lambda_\pm}{2})$. We have also used the notation $z_{ab} = z_a - z_b$.

{}From the Dirac equation, we can see that the second type of solution can
be obtained by replacing $(M,\psi_\pm)$ by $(-M,\mp \psi_\mp)$
as well as  $z$ with $\bar z$. This follows from the anti-automorphism obtained by composing \eqref{eq:antiautoonCA} with \eqref{eq:autoonC}. Explicitly, the solution is given by
\begin{align}
  &\psi_+  (\rho , z ) = \frac{-M+\frac12}{\pi} \int d^2 z ' e^{- \frac12 \rho}
 \left ( \frac{e^{-\rho} }{e^{-2 \rho} + |z - z '|^2}\right)^{-M+\frac32}
  \eta_+ (z ')  ~, \label{psiplusp}\\
  &  \psi_-  (\rho , z ) =  \frac{-M+\frac12}{\pi} \int d^2 z ' e^{ \frac12 \rho}
 \left ( \frac{e^{-\rho} }{e^{-2 \rho} + |z - z '|^2}\right)^{-M+\frac32}
  (\bar z- \bar z') \eta_+ (z ')  \label{psiminusp}
\end{align}
with the boundary behavior
\begin{align}
 \psi_+ (\rho , z ) \sim \eta_+ (z) e^{\rho (-M-1)} ~, \qquad \psi_- (\rho , z ) \sim
0~. \label{boundarycp}
\end{align}
{}From this solution, we can define another boundary operator with a different conformal weight.
We assign the boundary behavior by $\eta_+ (z) = \varepsilon_+ \delta^{(2)}(z-z_2)$.
Then from the asymptotic behavior around  $\rho \sim \infty$ and $z \neq z_2$, we can compute
the boundary two point function
\begin{align}
 \langle {\cal O}^{[+]}_F(z_1)   \widetilde{\cal O}^{[+]}_F (z_2) \rangle
 = - \frac{B^{[+]}_\psi (M - \frac12)}{\pi} \frac{1}{z_{12}^{2h} \bar z_{12}^{2 \bar h}}
 \label{2ptp}
\end{align}
 with dual conformal weight
$(h^{[+]} , \bar h^{[+]}) = (\frac{- M + 3/2}{2} , \frac{ - M + 1/2}{2}) = (\frac{1 + \lambda_\pm}{2} , \frac{ \lambda_\pm}{2})$.
It was proposed in \cite{CHR} that we should utilize the both types of boundary conditions
for the application to the AdS/CFT correspondence, see also \cite{CG}.

We can study the Dirac equation for $\psi_\pm = \tilde{C}^{3/2}_{\pm 1/2}$, which is given by
\eqref{diracp}, but with $M$ replaced by $-M$. Thus one type of solution is given by \eqref{psiplus}
and \eqref{psiminus}, but with $M$ replaced by $-M$. The conformal dimension of the dual operator is
$(\tilde h^{[-]} , \tilde{\bar h}^{[-]}) = (\frac{\lambda_\pm}{2}, \frac{1 + \lambda_\pm}{2})$. The second type is given by \eqref{psiplusp}
and \eqref{psiminusp} but with $M$ replaced by $-M$. The dual conformal dimension is
$(\tilde h^{[+]} , \tilde{\bar h}^{[+]}) = (\frac{2- \lambda_\pm}{2}, \frac{1 - \lambda_\pm}{2})$. This means that we have to have the opposite projection of $k$ on the tilded and untilded operators, as also seen from \eqref{eq:lagrangianstructure}.
In table \ref{tbl1}, the masses and the dual conformal dimensions are summarized.
As in \eqref{nulambda} we set $\lambda_+ = \lambda$ and $\lambda_- =  1-\lambda$ such that the AdS/CFT map becomes clear.
Notice that we can define two types of dual operators with conformal weights $(h^{[\pm]} , \bar h^{[\pm]})$ by changing the boundary conditions.%
\footnote{Precisely speaking, we construct two Dirac fermions by combining $C$ and $\tilde C$  as discussed at the end of section 4.1 of \cite{CHR}. We assign the different boundary condition for each Dirac fermion.}
  \begin{table}
   \begin{center}
    \begin{tabular}{|c|c|c|c|}
     \hline
      ~ & $(\text{mass})^2$ & $(h^{[+]},\bar h^{[+]}) $ & $ (h^{[-]},\bar h^{[-]}) $ \\ \hline
Scalar $(k=+1)$ & $- 1 + \lambda ^2$ & $(\frac{1+\lambda}{2},\frac{1+\lambda}{2} )$ &  $(\frac{1-\lambda}{2},\frac{1-\lambda}{2} )$ \\ \hline
Scalar $(k=-1)$ & $- 1 + (1-\lambda) ^2$ & $(\frac{2-\lambda}{2},\frac{2-\lambda}{2} )$ &  $(\frac{\lambda}{2},\frac{\lambda}{2} )$ \\ \hline
Spinor $(k=+1)$ & $ ( \lambda  - 1/2 )^2$ & $(\frac{1+\lambda}{2},\frac{\lambda}{2} ), (\frac{2-\lambda}{2},\frac{1-\lambda}{2} )$ &  $(\frac{1-\lambda}{2},\frac{2-\lambda}{2} ), (\frac{\lambda}{2},\frac{1+\lambda}{2}) $ \\ \hline
Spinor $(k=-1)$ & $ ( \lambda  - 1/2 )^2$ & $(\frac{2-\lambda}{2},\frac{1-\lambda}{2} ), (\frac{1+\lambda}{2},\frac{\lambda}{2} ) $  &   $(\frac{\lambda}{2},\frac{1+\lambda}{2} ),(\frac{1-\lambda}{2},\frac{2-\lambda}{2} )$ \\ \hline
    \end{tabular}
    \end{center}
  \caption{The masses of massive matters and the conformal weights of their dual operators
are summarized. Dual conformal weights for four complex massive scalars are all different and
those for Dirac fermions are divided into two classes.}\label{tbl1}
  \end{table}

\subsection{Three point function with a spin one current}
\label{abelian}

The main aim of this paper is to compute boundary three point functions of two fermionic operators
and a higher spin current with spin $s$. As a preparation, we compute the three point function
with a spin one  current inserted. Following the method in \cite{AKP}, we introduce the effect of such a U(1)
gauge field by a gauge transformation. This is possible since the bulk Chern-Simons gauge theory
has no dynamical fields.
The action of the U(1) Chern-Simons theory coupled to a
Dirac fermion is
\begin{align}
 S = \frac{k}{4\pi} \int A \wedge d A + \frac12 \int d^3 x \sqrt{g}
 (\bar \psi \not \!\! D \psi + M \bar \psi \psi )
\end{align}
with $D_\mu = \partial_\mu + A_\mu $. We study the first type of boundary conditions above for the fermions and demand the behaviour
at $\rho \to \infty$ to be
\begin{align}
 \hat \psi_+ \sim 0 ~, \qquad
\hat \psi_- \sim \varepsilon_- \delta^{(2)} (z - z_2)e^{ - \rho (1-M)} ~, \qquad
 \hat A \sim \mu \delta^{(2)} (z - z_3)
 \label{u1bc}
\end{align}
with a fermionic parameter $\varepsilon_-$. Then the three point function can be
found by examining the asymptotic behaviour of $\hat\psi_+$ and keeping only the term proportional to $\epsilon_-\mu$
\begin{align}
 {\cal O}(z_1) = \varepsilon_- \mu
\langle {\cal O}^{[-]}_F (z_1) \widetilde{\cal O}^{[-]}_F (z_2) J^{(1)} (z_3) \rangle + \cdots ~,
\qquad
 \hat \psi_+ (z) \sim \frac{{\cal O} (z) }{B^{[-]}_\psi} e^{ - \rho (1+M)}
\end{align}
around $\rho \to \infty$ and $z \neq z_2 , z_3$ as for the boundary two point function.
We can study the case with the second boundary condition in the same way.

We start from the free fermion with no U(1) gauge field i.e. $A=0$.
Then the three point function should be reduced to the two point function \eqref{2pt0}
with $(h , \bar h ) = (M + \frac12 , M + \frac32)$.
We introduce a non-zero gauge field with the boundary behavior \eqref{u1bc} by performing a
gauge transformation
\begin{align}
 A_\mu = \partial_\mu \Lambda ~, \qquad \Lambda (z) = \frac{\mu}{2\pi}
\frac{1 }{z-z_3} ~,
\end{align}
where we have used $\bar \partial z^{-1} = 2 \pi \delta^{(2)} (z) $.
The gauge transformation also acts on the fermions as
\begin{align}
 \psi_\pm (\rho , z ) \to \hat \psi_\pm = (1 - \Lambda (z)) \psi_\pm ~.
\end{align}
The boundary behavior around $\rho \to \infty$ should be
\begin{align}
 \hat \psi_- (\rho , z) \sim (1 - \Lambda (z)) \eta_- (z) e^{- \rho (1-M)}
  = \varepsilon_- \delta ^{(2)} (z - z_2) e^{- \rho (1-M)}
\end{align}
due to the boundary condition \eqref{u1bc}. This leads to
\begin{align}
 \eta _- (z) = \varepsilon_- (1 + \Lambda (z)) \delta ^{(2)} (z - z_2) ~.
\end{align}
{}From the asymptotic behavior of \eqref{psiplus} around $\rho \sim \infty, z \neq z_2,z_3$,
we find
\begin{align}
 {\cal O} (z_1) = - \varepsilon_-\mu\frac{(M+\frac12)B^{[-]}_\psi}{\pi}
  \left( \frac{\Lambda (z_2) - \Lambda (z_1) }{z_{12}^{M+\frac12} \bar z_{12}^{M+\frac32}}\right) +\ldots
\end{align}
thus giving
\begin{align}\label{}
    \langle {\cal O}^{[-]}_F (z_1) \widetilde{\cal O}^{[-]}_F (z_2) J^{(1)} (z_3) \rangle= \frac{1}{2\pi} \left( \frac{z_{12}}{z_{13} z_{23}} \right)
  \langle  {\cal O}^{[-]}_F (z_1) \widetilde  {\cal O}^{[-]}_F (z_2)  \rangle ~.
\end{align}
Here we note that the right hand side of the above equation
is the same as (4.13) of \cite{AKP} for the bosonic case.

\section{Correlation functions from the supergravity}
\label{3ptfn}

In this section, we compute boundary three point functions with two fermionic operators and
one higher spin current as in \eqref{3pti}
\begin{align}
 \langle {\cal O}_F^{(h , \bar h)} (z_1) \widetilde  {\cal O}_F^{(h , \bar h)} (z_2) J^{(s)\pm} (z_3) \rangle
\end{align}
from the supergravity theory of Prokushkin and Vasiliev \cite{PV1}.
We closely follow the method used for the $s = 1$ case in the previous section.
Namely, we introduce the effect of gauge field by making use of gauge transformations.
First we study how the higher spin gauge transformation acts on the massive fermions,
and then move to the computation of the three point functions.

\subsection{Higher spin gauge transformation}
\label{hsgtc}

In the previous section, we considered U(1) Chern-Simons theory coupled with massive fermions.
Now the theory is the one studied in section
\ref{vasiliev} and the field equations for the massive fermions are given in \eqref{feforc}.
The field equations are invariant under the following gauge transformation
\begin{align}
 &\delta A = d \Lambda + [A,\Lambda]_* ~, \qquad
 \delta \bar A = d \bar \Lambda + [\bar A, \bar \Lambda]_* ~,  \\
 &\delta C = C * \bar \Lambda - \Lambda * C ~, \qquad
 \delta \tilde C = \tilde C *  \Lambda - \bar \Lambda * \tilde C \label{gaugetrans} ~.
\end{align}
Since the transformation is much more complicated than that for the U(1) Chern-Simons theory,
we study it in more detail before applying it in the computation of boundary three point functions.

We would like to consider boundary three point functions with a higher spin current $J^{(s)\pm}(z_3)$.
The dual configuration of a gauge field in the bulk can be constructed by a gauge transformation with
a gauge parameter \cite{AKP}
\begin{align}
 \Lambda (\rho ,z ) = \sum_{n=1}^{2s-1} \frac{1}{(n-1)!} (- \partial)^{n-1}
  \Lambda^{(s)} (z) e^{(s-n)\rho} V^{(s)\pm}_{s-n} ~, \qquad \Lambda^{(s)} (z)
  = \frac{1}{2\pi} \frac{1}{z - z_3} ~,
  \label{lambdadef}
\end{align}
where the generators are defined in \eqref{bosong}.
In this paper we introduce bosonic higher spin fields and only discuss fermionic ones later.
The source term is the leading term in $A_{\bar z}$
\begin{align}\label{eq:changeAzbar}
 \delta A_{\bar z} = \partial_{\bar z} \Lambda^{(s)} e^{(s-1) \rho} V_{s-1}^{(s)\pm} + \cdots ~,
\end{align}
where the subleading terms are needed to satisfy the field equations \eqref{CSeq}.
The dual current $J^{(s)\pm}$ is in  $A_{z}$ as
\begin{align}\label{eq:changeAz}
 \delta A_{z} = \frac{1}{B^{(s)\pm}} J^{(s)\pm} e^{-(s-1) \rho} V_{-(s-1)}^{(s)\pm} ~, \qquad
   J^{(s)\pm} =  \frac{B^{(s)\pm}}{(2s-2)!} \partial^{2s-1}  \Lambda^{(s)}~.
\end{align}
Here $B^{(s)\pm}$ represents the coupling between the source and the dual current.

Since we introduce the gauge field by using a gauge transformation, we also need to know the transformation of the massive fields as in \eqref{gaugetrans}. Below we study the massive scalars first and then move
to the massive fermions.

\subsubsection{Gauge transformation for massive scalar fields}

As explained in section \ref{perturbations}, the massive fields are given by the mode expansions
of $C_\pm$ and $\tilde C_\pm$. The bosonic truncation can be done by restricting $s$ to be
integer. For simplicity we focus on $C=C_+$ and $J^{(s)}=J^{(s)+}$, but we can easily
generalize to the other cases. The scalar field corresponds to the first mode $C_0^1$ and its change
under the gauge transformation is
\begin{align}
 \hat C_0^1 = C_0^1 + (\delta C)_0^1 = C_0^1 - (\Lambda * C)_0^1 ~.
\end{align}
With the lone star product \eqref{lsp}, we can write the change explicitly as
\begin{align}
 (\delta C)_0^1 = - \sum_{n=1}^{2s - 1} \frac{1}{(n-1)!} (- \partial)^{n-1} \Lambda^{(s)}
\tfrac12 g_{2s-1}^{ss} (s-n , n-s) C_{n-s}^s e^{(s-n)\rho} ~.
\label{deltac01}
\end{align}
The main task here is to express $C_{n-s}^s$ in terms of the dynamical scalar field $C_0^1$.

Let us examine the field equations \eqref{fe1}, \eqref{fe2} and \eqref{fe3}.
If we set $m$ to the extremal value $m=-s+1$ i.e. $m=-|m|$ and $s=|m|+1$ in \eqref{fe2}, then the equation is simplified
since now only the first and the last terms remain. Solving the equation, we
find
\begin{align}
 C_{-|m|}^{|m|+1} = \left( \prod_{l=2}^{n+1} g^{2l}_3 (1,1-l)\right)^{-1} (-2 e^{- \rho} \partial_z)^{|m|} C_0^1 ~. \label{cmmp1}
\end{align}
In the same way, we obtain
\begin{align}
 C_{|m|}^{|m|+1} = \left( \prod_{l=2}^{n+1} g^{l2}_3 (l-1,-1)\right)^{-1} (2 e^{- \rho} \partial_{\bar z})^{|m|} C_0^1  \label{cmmp2}
\end{align}
by solving the equation \eqref{fe3} with $m=|m|$ and $s=|m|+1$.
The other equation \eqref{fe1} relates $C_m^s$ with fixed $m$.
In other words, we can reduce $C_{\pm |m|}^s$ to $C_{\pm |m|}^{|m|+1}$ utilizing the
equation \eqref{fe1}. Then, with the help of  \eqref{cmmp1} or \eqref{cmmp2},
the mode  $C_{\pm |m|}^s$ for all $s$ and $|m|$ can be written in terms of $C_0^1$.

The above argument actually applies both for integer and half integer $s$.
However, the equation \eqref{fe1} can be solved easier for integer $s$ since
$h_m^s = 0$ for the case, and indeed the solution was written as (4.42) in \cite{AKP}.
Using the solution, the gauge transformation was written as
\begin{align}\label{eq:changeCinfs}
& (\delta C)_0^1 = D^{(s)} C_0^1 ~, \qquad D^{(s)} =
 \sum_{n=1}^s f^{s,n} (\lambda , \partial_\rho) \partial^{n-1} \Lambda^{(s)} \partial^{s-n} ~.
\end{align}
One thing worth noting here is the upper bound in the sum over $n$. In the above
equation, $n$ is summed until $n=s$ while in \eqref{deltac01} it was until $n=2s-1$.
This is because for $n-s < 0$ there will be a factor $e^{-(s-n)\rho}$ due to \eqref{cmmp1} cancelling the factor $e^{(s-n)\rho}$ in \eqref{deltac01}. On the other hand for $n-s >0$ we have $e^{(s-n)\rho}$ due to \eqref{cmmp2} giving a total factor in \eqref{deltac01} of $e^{2(s-n)\rho}$ which vanishes in the large $\rho$
limit.

We need the explicit expression for $f^{s,n} (\lambda , \partial_\rho) $  when $\partial_\rho$
is replaced by $- (1 \pm \lambda)$. Denoting  $f^{s,n}_\pm (\lambda) = f^{s,n} (\lambda , - (1 \pm \lambda)) $,
it is given as (4.50) in \cite{AKP}:
\begin{align}
& f^{s,n}_\pm (\lambda )
 = (-1)^s \frac{\Gamma (s \pm \lambda)}{\Gamma (s-n+1 \pm \lambda)}
   \frac{1}{2^{n-1} (2[\frac{n}{2}] - 1)!! [\frac{n-1}{2}]!} \prod_{j=1}^{[\frac{n-1}{2}]}
   \frac{s+j-n}{2s-2j-1} ~.
   \label{fsn}
\end{align}

\subsubsection{Gauge transformation for massive spinor fields}

For  the massive fermions we again use the mode expansions of $C=C_+$.
Here we only consider bosonic gauge transformations,
and these relate half-integer spin fermionic modes to fermionic modes. We can thus make a fermionic truncation by restricting to $s \in \mathbb{Z} + 1/2$.
The massive fermion corresponds to $C^{3/2}_{\pm 1/2}$ and it shifts under the gauge transformation as
\begin{align}
 \hat C^{\frac32}_{\pm \frac12} = C^{\frac32}_{\pm \frac12} +
  (\delta C)^{\frac32}_{\pm \frac12} = C^{\frac32}_{\pm \frac12}
   - (\Lambda * C)^{\frac32}_{\pm \frac12} ~.
\end{align}
One way to obtain the explicit form of $(\delta C)^{\frac32}_{\pm \frac12}$ is
to solve the equations  \eqref{fe1}, \eqref{fe2} and \eqref{fe3}
directly as in the case with integer $s$.
But instead we would like to use a trick here.

One problem for the direct computation is that we do not know the explicit form of the star products
\eqref{lsp} with half-integer $s,t$ involved. Thus, it is convenient to define
the following fields by the action of $V^{3/2}_{\pm 1/2}$ from the right hand side as (using \eqref{eq:V32right})
\begin{align}\label{eq:fermiontrick}
 C^B_{(1)} \equiv C * V^{\frac32}_{\frac12}
 = \sum_{s,m}( C^B_{(1)})^{s}_{m} V^{~s}_m ~, \qquad
(C^B_{(1)})^{s}_{m} = C^{s-\frac12}_{m-\frac12}
 - \tfrac{(s-1-m)(2s+3-2\lambda)}{8(s-1)}  C^{s+\frac12}_{m-\frac12}  ~, \\
 C^B_{(2)} \equiv C * V^{\frac32}_{-\frac12}
 = \sum_{s,m} (C^B_{(2)})^{s}_{m} V^s_m ~, \qquad
(C^B_{(2)})^{s}_{m} = C^{s-\frac12}_{m+\frac12}
 + \tfrac{(s-1+m)(2s+3-2\lambda)}{8(s-1)}  C^{s+\frac12}_{m+\frac12}  ~.
\end{align}
Then, we can use the star product \eqref{lsp} with the known coefficients \eqref{gust}
as the index $s$ runs over integer values in terms of $C^B_{(1,2)}$.
Since
$(C_{(1,2)}^B)_0^1$ is proportional to $C^{3/2}_{\pm 1/2}$ as
\begin{align}
(C^B_{(1)})^{1}_{0} = - \tfrac12 (3-\lambda) C^{\frac32}_{-\frac12} ~, \qquad
(C^B_{(2)})^{1}_{0} = \tfrac12 (3-\lambda) C^{\frac32}_{\frac12} ~,
\end{align}
we can read off  $(\delta C)^{\frac32}_{\pm \frac12}$ from
\begin{align}
  (\delta  C^B_{(1,2)})^{1}_{0} =  - (\Lambda * C^B_{(1,2)})^{1}_{0} ~,
\end{align}
which can be obtained by multiplying $V_{\pm \frac12}^{\frac32}$ from the
right hand side of \eqref{gaugetrans}.
Using the lone star product \eqref{lsp} we have now
\begin{align}
 (\delta C_{(1,2)}^B)_0^1 = - \sum_{n=1}^{2s - 1} \frac{1}{(n-1)!} (- \partial)^{n-1} \Lambda^{(s)}
\tfrac12 g_{2s-1}^{ss} (s-n , n-s) (C_{(1,2)}^B)_{n-s}^s e^{(s-n)\rho} ~.
\end{align}
We again need to express $(C_{(1,2)}^B)_m^s$  in terms of $(C_{(1,2)}^B)_0^1$
via the field equations.

The field equations for $ C^B_{(1,2)}$ can be obtained by multiplying $V_{\pm \frac12}^{\frac32}$
from the right hand side of \eqref{feforc} as
\begin{align} \label{}
 &(d - \tfrac12 d \rho)  C^B_{(1)} + A * C^B_{(1)}  - C^B_{(1)}  * \bar A
 - e^{\rho} d \bar z C^B_{(2)} = 0 ~, \\
 &(d + \tfrac12 d \rho)  C^B_{(2)} + A * C^B_{(2)}  - C^B_{(2)}  * \bar A = 0 ~.
 \label{2ndeq}
\end{align}
In terms of the modes, we have
\begin{align}
 & (\partial_\rho - \tfrac12 ) (C^B_{(1)})^s_m + 2 (C^B_{(1)})^{s-1}_m + g^{(s+1)2}_3 (m,0) (C^B_{(1)})^{s+1}_m = 0 ~,
\nonumber \\
 & \partial (C^B_{(1)}) ^s_m + e^{\rho} [(C^B_{(1)}) ^{s-1}_{m-1} + \tfrac12 g^{2s}_2 (1,m-1) (C^B_{(1)}) ^s_{m-1}
 + \tfrac12 g^{2 (s+1)}_3 (1,m-1) C^B_{(1)}) ^{s+1}_{m-1} ] = 0 ~ , \nonumber \\
 & \bar \partial (C^B_{(1)}) ^s_m - e^\rho [(C^B_{(2)}) ^{s-1}_{m+1}  + \tfrac12 g^{s2}_2 ( m+1 , - 1) (C^B_{(1)}) ^s_{m+1} \nonumber \\
 & \qquad \qquad \qquad \qquad \qquad \qquad
 + \tfrac12 g^{ (s+1)2}_3 (m+1 , -1) (C^B_{(1)}) ^{s+1}_{m+1} + (C^B_{(2)}) ^s_m ] = 0 \nonumber~,
\end{align}
while for $C^{B}_{(2)}$ we can use the bosonic result just by replacing $\partial_\rho$
by $\partial_\rho + \frac12$.  For  $C^{B}_{(1)}$, we not only have the shift from $\partial_\rho$
to $\partial_\rho - \frac12$, but we  also have an effect from $C^{B}_{(2)}$.
Setting $s=m+1$, we get
\begin{align}
 \partial_{\bar z} (C^{B}_{(1)})_{m}^{m+1} - e^\rho [ \tfrac12 g^{ (s+1)2}_3 (m+1 , -1) (C^B_{(1)}) ^{m+2}_{m+1} + (C^B_{(2)}) ^{m+1}_m ] = 0 ~.
\end{align}
The solution to this equation is more complicated than \eqref{cmmp2}.
However, the above equation implies that $ (C^B_{(1)}) ^{m+1}_m \sim e^\rho (C^B_{(2)}) ^{m+1}_m
 \sim e^{\rho (1-m)} (C^B_{(2)}) ^{1}_0$ in the large $\rho$ limit up to the action of
$\partial_{\bar z}$.
This means that  only the contributions from $(C_{(1)}^B)_{n-s}^s$ with
$n-s \leq 0$ survives in the large $\rho$ limit.
 {}From this fact, we can safely neglect
the effects of $C^{B}_{(2)}$ in $C^{B}_{(1)}$.

{}From the above considerations, we conclude that
\begin{align}
 (\delta C)^{\frac32}_{\pm \frac12} = D_\pm^{(s)} C^{\frac32}_{\pm \frac12} ~, \qquad
 D_\pm^{(s)} = \sum_{n=1}^s f^{s,n} (\lambda, \partial_\rho \pm \tfrac12) \partial^{n-1} \Lambda^{(s)}
 \partial^{s-n} ~. \label{Ds}
\end{align}
When we can replace $\partial _ \rho \pm 1/2$ by $- (1 + \lambda)$ or $- (1 - \lambda)$,
the functions $f^{s,n} (\lambda, \partial_\rho \pm 1/2)$ become respectively  $f^{s,n}_+ (\lambda)$ or $f^{s,n}_- (\lambda)$  given in \eqref{fsn}.

\subsection{Three point functions with a generic spin current}

Now we have prepared for the computation of three point function
\eqref{3pti}
\begin{align}
 \langle {\cal O}_F^{(h , \bar h)} (z_1) \bar  {\cal O}_F^{(h , \bar h)} (z_2) J^{(s)\pm} (z_3) \rangle ~.
\end{align}
There are several kinds of correlators, but some of them can be obtained easily from others.
Here we only focus on $C=C_+$ but for $C_-$ we just need to replace $\lambda_+ = \lambda$
by $\lambda_- = 1 - \lambda$. We also consider only $J^{(s)} = J^{(s)+}$. The difference from
$J^{(s)-}$ is the multiplication of $k$ as in \eqref{bosong}. Since $k=\pm 1 $ when it acts on
the projected operator $C_\pm$, we just need to multiply the factor $k=\pm 1 $.
First we consider the operators with the conformal weight $(h , \bar h)= (\frac{1-\lambda}{2}, \frac{2-\lambda}{2})$, whose two point function has been computed in \eqref{2pt0}
in the holographic way. Secondly, we compute the case with $(h , \bar h)= (\frac{1+\lambda}{2}, \frac{\lambda}{2})$, which is dual to fermions associated to
the other boundary condition \eqref{boundarycp}. Finally we examine the case with
 $(h , \bar h)= (\frac{\lambda}{2}, \frac{1+\lambda}{2}),  (\frac{2- \lambda}{2}, \frac{1-\lambda}{2})$, which can be obtained by using the charge conjugated fields
$\tilde C$.

\subsubsection{An example}

We compute the three point function
\begin{align}
 \langle {\cal O}^{[-]}_F (z_1)   \widetilde{\cal O}^{[-]}_F (z_2)  J^{(s)} (z_3) \rangle  ~,
 \label{3ptF}
\end{align}
where ${\cal O}^{[-]}_F (z)$ has the conformal weight $(h , \bar h)= (\frac{1-\lambda}{2}, \frac{2-\lambda}{2})$. Setting the gauge field configuration $A=0$, the three point
function reduces to the two point function \eqref{2pt0}. As in the Abelian case  in subsection \ref{abelian}, we include the gauge field by utilizing the gauge transformation.

For  $A=0$, the solution for the
dual fermion is given by \eqref{psiplus}, \eqref{psiminus} with the asymptotic behavior
\eqref{boundaryc} around $\rho \sim \infty$. In this case we have $M=\frac12 - \lambda$.
We include a higher spin gauge field by the gauge transformation given in
\eqref{lambdadef}, which is a source to the higher spin current $J^{(s)}$ as discussed above.
The gauge transformation also changes the massive fermions as
\begin{align}
 \psi_\pm (\rho , z) \to \hat \psi_\pm (\rho , z) \sim (1 + D_\pm^{(s)}) \psi_\pm (\rho , z) ~,
\end{align}
where the differential operators are defined in \eqref{Ds}.
The asymptotic behavior of the fermion $\rho \sim \infty$ is
\begin{align}
 \hat \psi_+ (\rho ,z ) \sim 0 ~, \qquad
 \hat \psi_- (\rho , z)\sim (1 + D_-^{(s)}) e^{\rho (- \lambda - \frac12 )} \eta_- (z) ~.
\end{align}
In order to compute the boundary three point function \eqref{3ptF}, we need to assign the
boundary condition $\hat \psi_- (\rho , z) \sim \varepsilon_- e^{\rho (- \lambda - \frac12 )} \delta^{(2)} (z - z_2)$.
To linear order in the gauge transformation we thus have the relation
\begin{align}
 \eta_- (z) = \varepsilon_- (1 - D_-^{(s)}) \delta^{(2)} (z - z_2) ~, \qquad
 D_-^{(s)} = \sum_{n=1}^s f^{s,n}_+ (\lambda) \partial^{n-1} \Lambda^{(s)}
 \partial^{s-n} ~,
\end{align}
where $f^{s,n}_+ (\lambda)$ is defined in \eqref{fsn}.
Here we would like to remark that the coefficient  $f^{s,n}_+ (\lambda)$  becomes the same as the bosonic case due to the
shift from $\partial_\rho$ to $\partial_\rho - \frac12$.

The three point function \eqref{3ptF} can be now read off from the asymptotic behavior
of the massive fermion around  $\rho \sim \infty$, $z \neq z_2$.
From the asymptotic behavior $\psi_+ (\rho ,z) \propto e^{\rho (\lambda-\frac32)}$,
we find
\begin{align}
 \hat \psi_+ (\rho , z) \sim (1 + D^{(s)}_+) \psi_+ (\rho , z) ~, \qquad
 D_+^{(s)} = \sum_{n=1}^s f^{s,n}_- (\lambda) \partial^{n-1} \Lambda^{(s)}
 \partial^{s-n} ~.
\end{align}
Recall that there is a shift from $\partial_\rho$ to $\partial_\rho + \frac12$ in the
argument of $f^{s,n} (\lambda, \partial_\rho + \frac12)$ in \eqref{Ds}.
In terms of these differential operators and using \eqref{psiplus}, the three point function becomes
\begin{align}
 {\cal O} (z_1) = \frac{(\lambda - 1)B_\psi }{\pi}
 \left( D^{(s)}_+ (z_1) \frac{1}{z_{12}^{1 - \lambda} \bar z_{12}^{2 - \lambda}}
 - \int d ^2 z ' \frac{D_-^{(s)} (z ') \delta^{(2)} (z ' - z_2 ) }{(z_{1} - z ' )^{1 - \lambda}  (\bar z_{1} - \bar z ' )^{2 - \lambda}}  \right)\varepsilon_-+\ldots ~.
\end{align}
The bosonic counterpart is given by (4.28) of \cite{AKP}, and the only difference is that
our case has  $\bar z_{ab}^{2 - \lambda}$ while their case has $\bar z_{ab}^{1 - \lambda}$
(while we also need to exchange $\lambda$ by $- \lambda$).
Since the differential operators $D^{(s)}_\pm$ act on the holomorphic coordinate $z$,
the difference does not affect the result.
Therefore we can borrow their result and obtain
\begin{align}
\left \langle {\cal O}^{[-]}_F (z_1)  \widetilde {\cal O}^{[-]}_F (z_2 ) J^{(s)} (z_3) \right \rangle
 & = \frac{(-1)^{s-1} (\lambda - 1) B^{[-]}_\psi }
 {2 \pi^2 z_{12}^{1 - \lambda} \bar z_{12}^{2 - \lambda}}
 \frac{\Gamma (s)^2 \Gamma (s - \lambda)}{\Gamma (2s-1) \Gamma (1 - \lambda)}
 \left( \frac{z_{12}}{z_{13} z_{23}}\right)^s\\
 & =
 \frac{(-1)^{s-1}  }
 {2 \pi}
 \frac{\Gamma (s)^2 \Gamma (s - \lambda)}{\Gamma (2s-1) \Gamma (1 - \lambda)}
 \left( \frac{z_{12}}{z_{13} z_{23}}\right) ^s
 \left \langle {\cal O}^{[-]}_F (z_1)  \widetilde {\cal O}^{[-]}_F (z_2 )  \right \rangle \nonumber ~.
\end{align}
The result looks to be the same as (4.51) of \cite{AKP} for the bosonic case, but the middle
computation is different. There is the supersymmetry behind this fact as will be argued below.

\subsubsection{Alternative quantization}

In order to construct supergravity theory dual to the $\mathbb{C}$P$^N$ Kazama-Suzuki model,
we also need to assign the second type of boundary condition in \eqref{boundarycp}, as discussed in
\cite{CHR,CG}.
{}From the solution with the boundary condition given by \eqref{psiplusp}, \eqref{psiminusp},
we can compute the two point function for the dual operator $\mathcal{O}^{[+]}_F$
with $(h,\bar h) = (\frac{1+\lambda}{2},\frac{\lambda}{2})$ as \eqref{2ptp}.
The three point function
\begin{align}
\left \langle {\cal O}^{[+]} _F  (z_1)  \widetilde{\cal O}_F^{[+]}  (z_2 ) J^{(s)} (z_3) \right \rangle
\end{align}
can be then obtained by utilizing the gauge transformation as in the previous subsection.

The solution \eqref{psiplusp}, \eqref{psiminusp} is obtained by replacing
$(\frac12 - \lambda,\psi_\pm)$ by $(\lambda - \frac12,\mp\psi_\mp)$ along with $z$ by $\bar z$. {}Following the previous analysis, we then arrive at
\begin{align}
 &\left \langle {\cal O}_F^{[+]} (z_1) \widetilde {\cal O}_F ^{[+]} (z_2) J^{(s)} (z_3 ) \right \rangle  \\
 & \qquad =
  \frac{\lambda B^{[+]}_\psi }{\pi}
 \left( D^{(s)}_- (z_1) \frac{1}{z_{12}^{1+\lambda} \bar z_{12}^{\lambda}}
 - \int d ^2 z ' \frac{D_+^{(s)} (z ') \delta^{(2)} (z ' - z_2 ) }{(z_{1} - z ' )^{1+\lambda}  (\bar z_{1} - \bar z ' )^{\lambda}}  \right) ~, \nonumber
\end{align}
where the differential operators \eqref{Ds} are
\begin{align}
 D_\pm^{(s)} = \sum_{n=1}^s f^{s,n}_\mp (\lambda) \partial^{n-1} \Lambda^{(s)}
 \partial^{s-n} ~.
\end{align}
Again the differential operators act on the holomorphic coordinate $z$, and
 the bosonic result can be directly adopted. Thus, we find
\begin{align}
\left \langle {\cal O} ^{[+]} _F  (z_1)  \widetilde{\cal O}_F ^{[+]}  (z_2 ) J^{(s)} (z_3) \right \rangle
 & = \frac{(-1)^{s-1} \lambda B^{[+]}_\psi }
 {2 \pi^2 z_{12}^{1 + \lambda} \bar z_{12}^{\lambda}}
 \frac{\Gamma (s)^2 \Gamma (s + \lambda)}{\Gamma (2s-1) \Gamma (1 + \lambda)}
 \left( \frac{z_{12}}{z_{13} z_{23}}\right)^s\\
 & =
 \frac{(-1)^{s-1}  }
 {2 \pi}
 \frac{\Gamma (s)^2 \Gamma (s + \lambda)}{\Gamma (2s-1) \Gamma (1 + \lambda)}
 \left( \frac{z_{12}}{z_{13} z_{23}}\right) ^s
 \left \langle {\cal O} ^{[+]} _F  (z_1)   {\widetilde{\cal O}_F ^{[+]} } (z_2 )  \right \rangle \nonumber ~.
\end{align}

In summary, if we restore the choice of $k$-projection $\sigma=\pm$ on our dual operators ${\cal O} ^{[\delta]}_{F\sigma}$ we have obtained all the three-point functions with two fermionic matter fields and one bosonic higher spin current
\begin{multline}\label{eq:3pttotal}
\left \langle {\cal O} ^{[\delta]} _{F\sigma}  (z_1)  \widetilde{\cal O}_{F(-\sigma)}^{[\delta]}  (z_2 ) J^{(s)+} (z_3) \right \rangle
 \\
 =
 \frac{(-1)^{s-1}  }
 {2 \pi}
 \frac{\Gamma (s)^2 \Gamma (s +\delta \lambda_\sigma)}{\Gamma (2s-1) \Gamma (1 +\delta \lambda_\sigma)}
 \left( \frac{z_{12}}{z_{13} z_{23}}\right) ^s
 \left \langle {\cal O} ^{[\delta]} _{F\sigma}  (z_1)   {\widetilde{\cal O}_{F(-\sigma)} ^{[\delta]} } (z_2 )  \right \rangle  ~.
\end{multline}
Here it has been used that the tilded operator has the opposite $k$-projection, see \eqref{eq:lagrangianstructure}.

\subsubsection{Charge conjugation}

On the bulk side we can see what happens when we consider the gauge transformation on $\tilde C$ instead of $C$.  On the CFT side the dual field $\widetilde{\cal O}_{F}$ is obtained by charge conjugation.
We make use of the $\mathbb{Z}_4$ anti-automorphism \eqref{eq:antiautoonCA} which takes
\begin{align}
 \eta(C^{3/2}_{m,\sigma})=-i\tilde C^{3/2}_{m,-\sigma}\ ,\qquad \eta(\tilde C^{3/2}_{m,\sigma})=-i C^{3/2}_{m,-\sigma}\ ,\qquad \eta(A^s_{m})=(-1)^{-s} A^s_{m}\ .
\end{align}
Then, we see that for the correlators we get a factor $(-1)^s$ from $J^{(s)+}$ and an exchange of $k$-projection, i.e. using \eqref{eq:3pttotal}
\begin{multline}
\left \langle \widetilde{\cal O}_{F\sigma}^{[\delta]} (z_1 ){\cal O} ^{[\delta]} _{F(-\sigma)}  (z_2)  J^{(s)+} (z_3) \right \rangle
 \\
 =
 -\frac{1  }
 {2 \pi}
 \frac{\Gamma (s)^2 \Gamma (s +\delta \lambda_{-\sigma})}{\Gamma (2s-1) \Gamma (1 +\delta \lambda_{-\sigma})}
 \left( \frac{z_{12}}{z_{13} z_{23}}\right) ^s
 \left \langle {\widetilde{\cal O}_{F\sigma} ^{[\delta]} } (z_1 ){\cal O} ^{[\delta]} _{F(-\sigma)}  (z_2)   \right \rangle  ~.
\end{multline}
We can reproduce the same result by explicitly calculating the variation of  $\tilde C$ as mentioned above.
{}From the CFT side this result follows immediately by replacing $z_1$ and $z_2$ and changing the order of the fermionic operators on both sides.

\section{Bulk-boundary dictionary}
\label{dictionary}

In this section we will make the mapping of symmetries and states between bulk and boundary precise. This is done with a special focus on supersymmetry that we will use in the next section for calculations in the boundary CFT.

\subsection{Global transformations}

We can compare the global symmetries on both sides of the duality.
On the bulk side we find that the transformations that do not change the AdS$_3$ background solution \eqref{background} are of the form
\begin{align}
\begin{split}\label{eq:gaugetransrepeat}
 \Lambda^{\pm}_{s,m}&=\epsilon^\pm_{s,m}\sum_{m'=m}^{s-1}(-1)^{s-1-m'}\binom{s-1-m}{m'-m}z^{m'-m}e^{m'\rho}V_{m'}^{(s)\pm} \\
  &=\epsilon^\pm_{s,m}\sum_{n=1}^{2s-1} \frac{1}{(n-1)!} (- \partial)^{n-1}
  \Lambda^{(s)} (z) e^{(s-n)\rho} V^{(s)\pm}_{s-n} ~, \qquad \Lambda^{(s)} (z)
  = z^{s-1-m}\ .
\end{split}
\end{align}
As we know from eqs. \eqref{eq:changeAzbar}, \eqref{eq:changeAz}, this does not create any source current and is thus a global symmetry of the boundary CFT. Note that this works for both the bosonic and the fermionic case where $\epsilon^\pm_{s,m}$ is commuting or anti-commuting depending on the value of $s$. Using the automorphism relating $A$ and $\bar A$ (see above eq. \eqref{eq:autoonC}), we find the conjugated gauge transformations as well
\begin{align}
\begin{split}
 \bar\Lambda^{\pm}_{s,m}&=\bar\epsilon^\pm_{s,m}\sum_{m'=m}^{s-1}(-1)^{2s}\binom{s-1-m}{m'-m}z^{m'-m}e^{m'\rho}V_{-m'}^{(s)\pm} \\
  &=\bar\epsilon^\pm_{s,m}\sum_{n=1}^{2s-1} \frac{1}{(n-1)!} (-1)^{2s}  \partial ^{n-1}
  \bar\Lambda^{(s)} (z) e^{(s-n)\rho} V^{(s)\pm}_{-s+n} ~, \qquad \bar\Lambda^{(s)} (\bar z)
  = \bar z^{s-1-m}\ .
\end{split}
\end{align}

We postulate that the action on the dual fields are given by the OPE with
\begin{align}\label{eq:changeonbdry}
    \frac{1}{2\pi i}\oint dz \Lambda^{(s)} (z) J^{(s)\pm}(w)\ ,
\end{align}
where $J^{(s)\pm}$ are the dual currents with spin $s$. Let us consider an example. Using the bulk equations of motion and the asymptotic behavior, we find that the variation of $C^s_{m,\sigma}$ with respect to $\Lambda^+_{2,1}$ is
\begin{align}\label{}
    \delta_{\Lambda^+_{2,1}}C^s_{m,\sigma}=\del C^s_{m,\sigma} \ .
\end{align}
Remembering that the coupling to the boundary is of the form $\int d^2 z C|_{\textrm{bdry}} \mathcal{O}_C$, we see that $\mathcal{O}_C$ also has to transform like $\delta \mathcal{O}_C=\del\mathcal{O}_C=L_{-1}\mathcal{O}_C$. Note that the field $\mathcal{O}_C$ really is the dual to $\tilde C$ due to the conjugation in the Lagrangian \eqref{eq:lagrangianstructure}. On the CFT side the conjugation is the charge conjugation.
For the transformation corresponding to $\Lambda^+_{2,0}$ we get
\begin{align}\label{}
    \delta_{\Lambda^+_{2,0}}C^1_{0,\sigma}=-\frac12\del_\rho C^1_{0,\sigma} +z\del C^1_{0,\sigma} \ .
\end{align}
Using the asymptotic behavior and replacing $\del_\rho=2(h-1)$, we see that the boundary field has to transform as
\begin{align}\label{}
    \delta\mathcal{O}_{C^1_{0,\sigma}}=(h+z\del)\mathcal{O}_{C^1_{0,\sigma}} \ .
\end{align}
Which fits with the proposal giving $L_0\mathcal{O}_{C^1_{0,\sigma}}=h\mathcal{O}_{C^1_{0,\sigma}}$. This also works for $C^{3/2}_{\pm1/2}$.
Finally for $\Lambda^+_{s,m}$ with $m$ positive, we see from \eqref{eq:changeCinfs} that if we put the field at $z=0$ the dual boundary field will not transform, i.e. $L_1\mathcal{O}_{C^1_{0,\sigma}}=0$.

The leading term in the gauge transformation $\Lambda^{\pm}_{s,m}$ is $(-1)^{(s-1-m)}e^{m\rho}V^{(s)\pm}_m$ whose dual under the automorphism above \eqref{eq:autoonC} simply is $V^{(s)\pm}_{-m}$. We see that it is natural that $\Lambda^{+}_{2,m}$ is related to $L_{-m}$.
Indeed we find that the following identification fulfill the global part of the superconformal algebra \eqref{eq:superconfalg}
\begin{align}\label{eq:superconformaldict}
    L_m\leftrightarrow (-1)^{m+1}V^{2+}_{-m}\ ,\qquad U_0\leftrightarrow\frac{\nu+k}{2}\ ,\qquad G^\pm_m\leftrightarrow
    (-1)^{m+1/2}\sqrt{2}P_\pm V_{-m}^{3/2+} \ .
\end{align}
Explicitly the transformations related to the supersymmetry transformations are
\begin{align}\label{eq:bulksusy}
    G^\pm_{-1/2} \leftrightarrow \Lambda^\pm=\epsilon^\pm\sqrt{2}P_\pm V^{(3/2)+}_{1/2}e^{\rho/2}\ ,\qquad \bar G^\pm_{-1/2} \leftrightarrow \bar\Lambda^\pm=-\bar\epsilon^\pm\sqrt{2}P_\pm V^{(3/2)+}_{-1/2}e^{\rho/2} \ .
\end{align}

\subsection{Currents}

If we extend the use of \eqref{eq:changeonbdry} for $\Lambda^{(s)} (z)=(z-w)^{-1}$, we create insertions of the current $J^{(s)\pm}(w)$. As in \cite{CY} we split the gauge field up into the AdS$_3$ part
 $A_\text{AdS}$ in \eqref{background} and the small deformation $\Omega$ as
\begin{align}
 A=A_\text{AdS}+\Omega\ .
\end{align}
The linearized equation of motion for $\Omega$ is
\begin{align}\label{eq:eqmomega}
 d\Omega+A_\text{AdS}\wedge_*\Omega+\Omega\wedge_*A_\text{AdS}=0\ ,
\end{align}
and the needed extra boundary action is
\begin{align}
 S_{\textrm{bdry}}=-\int d^2 z e^{2\rho}\str(\Omega_z\Omega_{\bar z}) \ .
\end{align}

On the bulk side, when we deform the AdS$_3$ gauge field using \eqref{eq:gaugetransrepeat} with general $\Lambda^{(s)}$ we get a solution to the equations of motion
\begin{align}\label{eq:omegasol}
    \Omega^{(s)\pm}_z&=\epsilon\frac{1}{(2s-2)!}\del^{2s-1}\Lambda^{(s)}(z)e^{-(s-1)\rho}V^{(s)\pm}_{-(s-1)}\ ,\\
    \Omega^{(s)\pm}_{\bar z}&=\epsilon\sum_{n=1}^{2s-1} \frac{1}{(n-1)!} (- \partial)^{n-1}
  \bar\del\Lambda^{(s)} (z) e^{(s-n)\rho} V^{(s)\pm}_{s-n}\sim 2\pi\delta^{(2)}(z-w)e^{(s-1)\rho}V^{(s)\pm}_{s-1}+\ldots\ , \nonumber \\
    \Omega^{(s)\pm}_\rho&=0\ . \nonumber
\end{align}
See eqs. \eqref{lambdadef}, \eqref{eq:changeAzbar} and \eqref{eq:changeAz} above.
We only need to remember the source term in $\Omega_{\bar z}$ which is the leading term in the $\rho$-expansion. The remaining terms are fixed by the equations of motion, given the form of $\Omega_z$.
It is nicer to write the field $\Omega$ out into components $\Omega=\sum_{s,m,\sigma}\Omega^{(s)\sigma}_{m}V^{(s)\sigma}_m$, and then define the coupling to the boundary current as
\begin{align}\label{}
    \exp\big(-\frac{1}{2\pi}\int d^2z [(\Omega_{\bar z})^{(s)\sigma}_{s-1}]|_{\textrm{bdry}}J^{(s)\sigma}\big)\ .
\end{align}
This means that $J^{(s)\sigma}$ has conformal weight $s$. Here we have a factor of $2\pi$ compared to earlier sections in the bulk-boundary couplings to be in harmony with eq. \eqref{eq:changeonbdry}.

We can now find the changes under the supersymmetry algebra using the equations of motion for $\Omega^{(s)\sigma}_{m}$ found via \eqref{eq:eqmomega}. This determines the supersymmetry structure on the CFT side. We expect the higher spin currents to organize in multiplets $W^{s0},W^{s\pm},W^{s1}$, see appendix \ref{CFTW}. We readily fix the correspondence for the lowest supermultiplet -- the superconformal algebra -- using the result of the last subsection:
\begin{align}\label{}
\begin{split}
     W^{11}&\leftrightarrow \Omega^{2+} \sim V^{2+}_1\ ,\qquad W^{10}\leftrightarrow \frac\nu2\Omega^{1+}+\frac{1}2\Omega^{1-}\sim\frac{\nu+k}{2}\ ,\\ W^{1\pm}&\leftrightarrow \frac{\sqrt{2}}2(\Omega^{3/2+}\pm \Omega^{3/2-})\sim\sqrt{2}P_\pm V_{1/2}^{3/2+}\ .
\end{split}
\end{align}
Here the similarity sign is just the mnemonic rule for the generator in the leading term. In the general case we need the dual of $W^{s1}$ to be independent of $k$, otherwise $G^\pm_{-1/2}$ will give higher spin solutions. We thus fix the normalization as $W^{s1}\leftrightarrow \Omega^{(s+1)+}$, and we then obtain the rest by working with the duals of $G^\pm_{-1/2}$ in \eqref{eq:bulksusy} and comparing with \eqref{eq:susytranswalg} as
\begin{align}\label{}
\begin{split}
     W^{s1}&\leftrightarrow \Omega^{(s+1)+} \sim V^{(s+1)+}_s\ ,\qquad W^{s0}\leftrightarrow \frac{\nu+(2s-1)k}{4(s-1/2)}\Omega^{(s)+}\sim\frac{\nu+(2s-1)k}{4(s-1/2)}V^{(s)+}_{s-1}\ ,\\ W^{s\pm}&\leftrightarrow \frac{\sqrt{2}}2(\Omega^{(s+1/2)+}\pm \Omega^{(s+1/2)-})\sim\sqrt{2}P_\pm V_{s-1/2}^{(s+1/2)+}\ ,
\end{split}
\end{align}
or in terms of the currents $J^{(s)\pm}$
\begin{align}\label{eq:cftcurrentsinbulknota}
\begin{split}
 W^{s0}&=\frac{\nu}{4(s-1/2)}J^{(s)+}+\frac12J^{(s)-}\ ,\\
 W^{s\pm}&=\frac1{\sqrt{2}}(J^{(s+1/2)+}\pm J^{(s+1/2)-})\ ,\\
 W^{s1}&=J^{(s+1)+}\ .
\end{split}
\end{align}

\subsection{States}

Finally, we can discuss how the boundary states should transform given the knowledge from the bulk side. We will denote the solutions to the equations of motion by
\begin{align}\label{}
    C^{1[\delta]}_{0\pm}\sim \phi_0\delta^{(2)}(z-w)e^{(-1+\delta\lambda_\pm)\rho}\ ,
\end{align}
where, as above, we denote standard/alternate quantization by $\delta=\pm$, and we just show the lowest component of the full solution.
Indeed for $\lambda$ positive, the standard quantization leads to the asymptotically fastest growing solution. The dual operators will have conformal weights
\begin{align}\label{eq:genbosweights}
 h^{[\delta]}_{\pm}=(1+\delta\lambda_\pm)/2\ .
\end{align}
For the fermions we name the boundary conditions by
\begin{align}\label{}
    C^{3/2[\delta]}_{\delta/2\pm}&\sim \eta_\delta\delta^{(2)}(z-w)e^{(-1-\delta/2+\delta\lambda_\pm)\rho}\ , & C^{3/2[\delta]}_{-\delta/2\pm}&\sim 0\ ,
\end{align}
where the conformal weights of the dual operators are
\begin{align}\label{eq:genfermweights}
 h^{[\delta]}_{\pm} &= \frac{1+\delta\lambda_\pm}2\ ,  & \bar h^{[\delta]}_{\pm}&=\frac{(1-\delta 1 )+\delta\lambda_\pm}2\ .
\end{align}

The coupling to the boundary fields is (suppressing coupling constants)
\begin{align}\label{}
    \int d^2 z \sum_{\sigma=\pm,\delta=\pm1}\big(C^{1[\delta]}_{0\sigma}|_{\textrm{bdry}}\mathcal{O}^{1[\delta]}_{0\sigma}+
    C^{3/2[\delta]}_{\delta/2\sigma}|_{\textrm{bdry}}\mathcal{O}^{3/2[\delta]}_{\delta/2\sigma}\big) \ ,
\end{align}
and from this we can find the supersymmetry transformation of the boundary fields by using \eqref{eq:bulksusy} on the bulk fields. The important relations are
\begin{align}\label{}
    \delta_{\Lambda^{\pm}}\mathcal{O}^{1[-]}_{0\pm}&=0\ , & \delta_{\Lambda^{\pm}}\mathcal{O}^{3/2[-]}_{-1/2\pm}&=0\ ,\\
    \delta_{\Lambda^{\pm}}\mathcal{O}^{1[-]}_{0\mp}&=\epsilon^{\pm}2\sqrt{2}\frac{\lambda_\pm-1}{\lambda_\pm}\mathcal{O}^{3/2[+]}_{1/2\pm}  \ , & \delta_{\Lambda^{\pm}}\mathcal{O}^{3/2[-]}_{-1/2\mp}&=\epsilon^\pm\frac{1-\lambda_\pm}{\sqrt{2}}\mathcal{O}^{1[+]}_{0\pm}\ ,
\end{align}
and for the anti-chiral transformations (via conjugation)
\begin{align}\label{}
    \delta_{\bar\Lambda^{\pm}}\mathcal{O}^{1[-]}_{0\mp}&=0\ , & \delta_{\bar\Lambda^{\pm}}\mathcal{O}^{3/2[+]}_{1/2\pm}&=0\ ,\\
    \delta_{\bar\Lambda^{\pm}}\mathcal{O}^{1[-]}_{0\pm}&=\bar\epsilon^{\pm}2\sqrt{2}\frac{\lambda_\pm}{\lambda_\pm-1}\mathcal{O}^{3/2[-]}_{-1/2\pm}  \ , & \delta_{\bar\Lambda^{\pm}}\mathcal{O}^{3/2[+]}_{1/2\mp}&=-\bar\epsilon^\pm\frac{1-\lambda_\mp}{\sqrt{2}}\mathcal{O}^{1[+]}_{0\mp}\  .
\end{align}
Here we note that $\Lambda^\pm$ changes sign on $k$ and hence also choice of boundary conditions:
\begin{align}\label{}
\begin{array}{ccccc}
    &   & \mathcal{O}^{1[-]}_{0\sigma} &  & \\
   & ^{\Lambda^{-\sigma}}\swarrow &  & \searrow^{\bar\Lambda^\sigma} & \\
   \mathcal{O}^{3/2[+]}_{1/2-\sigma}&  &  &  & \mathcal{O}^{3/2[-]}_{-1/2\sigma} \\
   & _{\bar\Lambda^\sigma}\searrow &  & \swarrow_{\Lambda^{-\sigma}} &  \\
   &  & \mathcal{O}^{1[+]}_{0-\sigma} &  &
\end{array}
\end{align}
For the conjugated fields we obtain
\begin{align}\label{}
    \delta_{\Lambda^{\pm}}\mathcal{\widetilde{O}}^{1[-]}_{0\mp}&=0\ , & \delta_{\Lambda^{\pm}}\mathcal{\widetilde{O}}^{3/2[-]}_{-1/2\pm}&=0\ ,\\
    \delta_{\Lambda^{\pm}}\mathcal{\widetilde{O}}^{1[-]}_{0\pm}&=-\epsilon^{\pm}2\sqrt{2}\frac{\lambda_\pm}{\lambda_\pm-1}\mathcal{\widetilde{O}}^{3/2[+]}_{1/2\pm}  \ , & \delta_{\Lambda^{\pm}}\mathcal{\widetilde{O}}^{3/2[-]}_{-1/2\mp}&=-\epsilon^\pm\frac{1-\lambda_\mp}{\sqrt{2}}\mathcal{\widetilde{O}}^{1[+]}_{0\mp}\ ,\\
    \delta_{\bar\Lambda^{\pm}}\mathcal{\widetilde{O}}^{1[-]}_{0\pm}&=0\ , & \delta_{\bar\Lambda^{\pm}}\mathcal{\widetilde{O}}^{3/2[+]}_{1/2\pm}&=0\ ,\\
    \delta_{\bar\Lambda^{\pm}}\mathcal{\widetilde{O}}^{1[-]}_{0\mp}&=-\bar\epsilon^{\pm}2\sqrt{2}\frac{\lambda_\pm-1}{\lambda_\pm}\mathcal{\widetilde{O}}^{3/2[-]}_{-1/2\pm}  \ , & \delta_{\bar\Lambda^{\pm}}\mathcal{\widetilde{O}}^{3/2[+]}_{1/2\mp}&=\bar\epsilon^\pm\frac{1-\lambda_\pm}{\sqrt{2}}\mathcal{\widetilde{O}}^{1[+]}_{0\pm}\ ,
\end{align}
\begin{align}\label{}
\begin{array}{ccccc}
    &   & \mathcal{\widetilde{O}}^{1[-]}_{0\sigma} &  & \\
   & ^{\Lambda^{\sigma}}\swarrow &  & \searrow^{\bar\Lambda^{-\sigma}} & \\
   \mathcal{\widetilde{O}}^{3/2[+]}_{1/2\sigma}&  &  &  & \mathcal{\widetilde{O}}^{3/2[-]}_{-1/2-\sigma} \\
   & _{\bar\Lambda^{-\sigma}}\searrow &  & \swarrow_{\Lambda^{\sigma}} &  \\
   &  & \mathcal{\widetilde{O}}^{1[+]}_{0-\sigma} &  &
\end{array}
\end{align}
where we remember that the conjugated fermions have opposite mass, i.e. $\mathcal{\widetilde{O}}^{3/2[\delta]}_{\delta 1/2\sigma}$ has the same mass as $\mathcal{O}^{3/2[\delta]}_{\delta 1/2-\sigma}$.
Since we have a complex algebra, we can have two oppositely quantized solutions for each field.

\section{Comparison with dual $\mathbb{C}$P$^N$ model}
\label{CFT}

In \cite{CHR} we have proposed that the  higher spin ${\cal N}=2$ supergravity of Prokushkin
and Vasiliev \cite{PV1} is dual to a large $N$ limit of the ${\cal N}=(2,2)$
$\mathbb C$P$^N$ Kazama-Suzuki
model \eqref{sminimal}
\begin{align} \label{}
 \frac{\widehat{\text{su}}(N+1)_k \oplus \widehat{\text{so}}(2N)_1 }
{\widehat{\text{su}}(N)_{k+1}  \oplus \widehat{\text{u}}(1)_{N(N+1)(k+N+1)}}
\end{align}
with the combination
$
 \lambda = N /(k+N)
$
kept fixed. There is strong evidence supporting our claim, as mentioned in the introduction,
and we now want to provide further evidence by explaining the results for the correlators from the CFT side.

Before going into the details of the dual CFT analysis, let us first summarize the results
obtained from the supergravity side in section \ref{3ptfn}.
As discussed in section \ref{vasiliev}, the supergravity theory consists of higher spin
gauge fields and massive matters. There are bosonic and fermionic higher spin
gauge fields, but we have only considered bosonic fields generated by
 $V^s_m = V^{(s)+}_m$, whose dual current is denoted by $J^{(s)}(z)$.
For the massive matter,
the conformal weights of the dual operators
are summarized in table \ref{tbl1}.
The dual operators can be expressed by ${\cal O}_B^{(h, h)}$ for the bosonic ones
and  ${\cal O}_F^{(h,\bar h)}$ for the fermionic ones where
$(h,\bar h)$ denotes their conformal weights.

In the bosonic case, the three point function with one higher spin current and two
massive scalars has been computed in \cite{CY,AKP} as
\begin{align}
\label{bosonresult}
\left \langle  {\mathcal O}_B^{(h,h)}(z_1)  \widetilde  {\mathcal O}_B^{(h,h)}(z_2 ) J^{(s)} (z_3) \right \rangle
  = N_s (h)
 \left( \frac{z_{12}}{z_{13} z_{23}}\right) ^s
 \left \langle   {\mathcal O}_B^{(h,h)}(z_1)  \widetilde  {\mathcal O}_B^{(h,h)}(z_2 ) \right \rangle ~,
\end{align}
where the factor is given by
\begin{align}
N_s (h) =
 \frac{(-1)^{s-1}  }
 {2 \pi}
 \frac{\Gamma (s)^2 \Gamma (s - 1 + 2 h )}{\Gamma (2s-1) \Gamma (2 h )} ~.
 \label{Nsh}
\end{align}
For the complex dual operators, we have to multiply a factor $(-1)^s$.
In \cite{CY,AKP}, they only considered the $k=1$ sector with $h=(1 \pm \lambda)/2$, but
it is easy to extend to the $k=-1$ sector with $h=\lambda/2, (2 -\lambda)/2$.
In section \ref{3ptfn}, we have extended the computation to the case with fermionic
operators and the results \eqref{eq:3pttotal} can via \eqref{eq:genbosweights} and \eqref{eq:genfermweights} be summarized as
\begin{align}
\left \langle  {\mathcal O}_F^{(h,\bar h)}(z_1)  \widetilde  {\mathcal O}_F^{(h,\bar h)}(z_2 ) J^{(s)} (z_3) \right \rangle
  = N_s(h)
 \left( \frac{z_{12}}{z_{13} z_{23}}\right) ^s
 \left \langle   {\mathcal O}_F^{(h,\bar h)}(z_1)  \widetilde  {\mathcal O}_F^{(h,\bar h)}(z_2 ) \right \rangle  \label{sugraresult}
\end{align}
up to a phase factor $(-1)^s$.
Here $\bar h = h \pm 1/2$.
Notice that the factor $N_s(h)$ is the same as in the bosonic case.
In the rest of this section, we would like to explain the result \eqref{sugraresult} from the dual CFT viewpoint.

\subsection{Dual $\mathbb{C}$P$^N$ model}

We would like to explain the results \eqref{sugraresult} by considering how the map works between
the massive matter in the bulk and the dual operators at the boundary.
The Kazama-Suzuki model has a factorization of chiral and anti-chiral sectors.
Let us first focus on the chiral part. Then the primary states are labeled by the
representations of groups in the cosets as $(\rho,s;\nu,m)$.
The labels $\rho, \nu$ are highest weights of su$(N+1)$ and su$(N)$ and
the other labels $s,m$ are related to so$(2N)$ and u(1). As explained in \cite{CG}
the label $m$ is uniquely fixed by the other labels in the large $N$ limit, so it
will be suppressed in the following. Since we consider the NS-sector, we either
have the identity representation $(s=0)$ or the vector representation $(s=2)$ for so$(2N)$.
The conformal weights for the relevant states are \cite{CHR}
\begin{align}
 h (\text{f},s;0) = \frac{ \frac{s}{2} + \lambda}{2} ~, \qquad
 h (0,s;\text{f}) =  \frac{ 2 - \frac{s}{2} - \lambda}{2}
\end{align}
in the 't Hooft large $N$ limit. Here f denotes the fundamental representation, and
the conjugate operators are given by replacing f by the anti-fundamental representation
$\bar{\text{f}}$.

The states of the full CFT have labels both of
the chiral and the anti-chiral sectors.
The CFT partition function is of the form
\begin{align}
 Z (q) = |q^{-\frac{c}{24}}|^2 \sum_{\rho , \nu } \sum_{s , \bar s = 0 ,2 }
 b_{(\rho , \nu ; s)} (q) b_{(\rho , \nu ; \bar s)} (\bar q)  ~,
\end{align}
where $b_{(\rho , \nu ;s)} (q)$ is the branching function of the state $(\rho , \nu ;s)$.
One point here is that the NS-sector is given by the sum of $s=0$ and $s=2$ states.
Thus the states dual to the bosonic matter can be expressed as (simply identifying via the conformal weights \eqref{eq:genbosweights})
\begin{align}\label{eq:cosetfieldid}
 & \mathcal{O}^{1[-]}_{0-}=| \text{f},0;0 \rangle_L \otimes  |\text{f},0;0 \rangle_R~, \qquad
 \mathcal{O}^{1[+]}_{0+}=|\text{f},2;0\rangle_L \otimes  |\text{f},2;0\rangle_R ~,  \\
 & \mathcal{O}^{1[+]}_{0-}=|0,0;\text{f}\rangle_L \otimes |0,0;\text{f}\rangle_R  ~, \qquad
 \mathcal{O}^{1[-]}_{0+}=|0,2;\text{f}\rangle_L \otimes |0,2;\text{f}\rangle_R  ~,
\nonumber
\end{align}
and those dual to the fermionic matter are
\begin{align}
 & \mathcal{O}^{3/2[-]}_{-1/2-}=|\text{f},0;0\rangle_L \otimes  |\text{f},2;0\rangle_R ~, \qquad
 \mathcal{O}^{3/2[+]}_{1/2+}=|\text{f},2;0\rangle_L \otimes  |\text{f},0;0) _R~,  \\
 & \mathcal{O}^{3/2[+]}_{1/2-}=|0,0;\text{f}\rangle_L \otimes |0,2;\text{f}\rangle_R ~, \qquad
 \mathcal{O}^{3/2[-]}_{-1/2+}=|0,2;\text{f}\rangle_L \otimes |0,0;\text{f}\rangle_R  ~.
 \nonumber
\end{align}
The conformal weights of these states are the same as in table \ref{tbl1}.

As we saw in section \ref{vasiliev}, the generators $P_{\pm}V_m^{(s)+}$ with $s=2,3,\ldots$ generate hs$[\lambda_\pm]$. It is also known \cite{PRS} that the algebra can be realized as the quotient of the universal enveloping algebra $U(\text{sl}(2))$ by the ideal generated by fixing the quadratic Casimir to $(\lambda^2-1)/4$
\begin{align}\label{}
    \text{hs}[\lambda_\pm]\oplus\mathbb{C}=\frac{U(\text{sl}(2))}{\langle C_2-(\lambda^2-1)/4\rangle} \ .
\end{align}
In eq. \eqref{eq:superconformaldict} we saw that the dual action of $P_{\pm}V_m^{(2)+}$ on the states ${O}^{1[\delta]}_{0\pm}$ is given by $L_m$, and indeed we find that the quadratic Casimir, when acting on these states, has just the right value
\begin{align}\label{}
    C_2|{O}^{1[\delta]}_{0\pm}\rangle=(L_0^2-\frac12(L_{+1}L_{-1}+L_{-1}L_{+1}))|{O}^{1[\delta]}_{0\pm}\rangle=\frac14(\lambda^2_\pm-1)|{O}^{1[\delta]}_{0\pm}\rangle \ .
\end{align}
This now gives a representation of the higher spin algebra on our states which we identify with $(J_m^{(s)+}\pm J_m^{(s)-})/2$, where $m$ are the modes having $|m|<s$. In particular, we can find the action of the zero modes on our states which directly determines the pre-factor $N_s(h)$ in the three-point function \eqref{bosonresult}. The eigenvalue depends only on $\lambda$ and the conformal dimension of the state. For the projection onto $k=1$ we can thus directly take over the result of the analysis of the bosonic case made in \cite{AKP}.
\begin{align}
 & J_0^{(s)}  ( |\text{f},2;0\rangle_L \otimes  |\text{f},2;0\rangle_R  )
 = N_s (\tfrac{1+\lambda}{2} )  ( |\text{f},2;0\rangle_L \otimes  |\text{f},2;0\rangle_R )  ~, \\
 & J_0^{(s)}  ( |0,2;\text{f}\rangle_L \otimes |0,2;\text{f}\rangle_R  )
 = N_s (\tfrac{1-\lambda}{2} ) ( |0,2;\text{f}\rangle_L \otimes |0,2;\text{f}\rangle_R )  ~,
 \nonumber
\end{align}
where the coefficient $N_s(h)$ is defined in \eqref{Nsh}.
Replacing $\lambda$ by $1-\lambda $, we can also find
\begin{align}
 & J_0^{(s)}   ( |\text{f},0;0\rangle_L \otimes  |\text{f},0;0\rangle_R  )
 = N_s (\tfrac{\lambda}{2} ) ( |\text{f},0;0\rangle_L \otimes  |\text{f},0;0\rangle_R )  ~, \\
 & J_0^{(s)}  ( |0,0;\text{f}\rangle_L \otimes |0,0;\text{f}\rangle_R  )
 = N_s (\tfrac{2-\lambda}{2} )  ( |0,0;\text{f}\rangle_L \otimes |0,0;\text{f}\rangle_R )  ~.
 \nonumber
\end{align}
Now the point is that the higher spin generator $V_0^{(s)+}$ acts only on the chiral (left-moving) part,
so the argument immediately extends to the fermionic states. Namely, we obtain
\begin{align}
 &J_0^{(s)}   ( |\text{f},2;0\rangle_L \otimes  |\text{f},0;0\rangle_R  )
 = N_s (\tfrac{1+\lambda}{2} )  ( |\text{f},2;0\rangle_L \otimes  |\text{f},0;0\rangle_R )  ~,  \\
& J_0^{(s)}  ( |0,2;\text{f}\rangle_L \otimes |0,0;\text{f}\rangle_R  )
 = N_s (\tfrac{1-\lambda}{2} )  ( |0,2;\text{f}\rangle_L \otimes |0,0;\text{f}\rangle_R )  ~,
 \nonumber \\
 & J_0^{(s)}   ( |\text{f},0;0\rangle_L \otimes  |\text{f},2;0\rangle_R  )
 = N_s (\tfrac{\lambda}{2} ) ( |\text{f},0;0\rangle_L \otimes  |\text{f},2;0\rangle_R )  ~, \nonumber \\
 & J_0^{(s)}   ( |0,0;\text{f}\rangle_L \otimes |0,2;\text{f}\rangle_R  )
 = N_s (\tfrac{2-\lambda}{2} ) ( |0,0;\text{f}\rangle_L \otimes |0,2;\text{f}\rangle_R )  ~. \nonumber
\end{align}
This reproduces the supergravity results in \eqref{sugraresult}.

In principle we could also have used that the superalgebra shs$[\lambda]$ is generated by the enveloping algebra of the $\mathcal{N}=1$ superalgebra osp$(1|2)$ given in \eqref{eq:N1algebra}
\begin{align}\label{}
    \text{shs}[\lambda]\oplus\mathbb{C}=\frac{U(\textrm{osp}(1|2))}{\langle C_2-\lambda(\lambda-1)/4\rangle} \ ,
\end{align}
where $C_2$ is the quadratic Casimir of osp$(1|2)$. Instead, we will in the next section directly use the supersymmetry of the dual CFT to reproduce the results.

\subsection{$\mathcal{N}=(2,2)$ supersymmetry}

\label{susy}

We will now use the $\mathcal{N}=(2,2)$ supersymmetry of the dual CFT to reproduce the results from the bulk.

\subsubsection*{Two-point functions}

In the large $N$ limit we know that the coset fields in eq. \eqref{eq:cosetfieldid} are (anti-)chiral primaries \cite{CHR}, see also \cite{Candu:2012tr}. These fields come together with the fields built of anti-fundamental representations, and which have opposite supersymmetric chirality. On the bulk side these fields correspond to the tilded operators.

We will now switch to standard supersymmetry notation. In the superconformal theory we thus have two chiral fields which we denote $\phi_{h_\pm}$, where $h_\pm=(1-\lambda_\pm)/2$ is the conformal weight. Relating back to the bulk side notation we thus have
\begin{align}\label{}
    \phi_{h_+}&=\mathcal{O}^{1[-]}_{0+} \ , & \phi_{h_-}&=\mathcal{\widetilde{O}}^{1[-]}_{0-}\ .
\end{align}
The remaining fields in the supermultiplet we denote as (see appendix \ref{CFTW})
\begin{align}\label{}
    \psi_{h_\pm}&=G^-_{-1/2}\phi_{h_\pm}\ , & \bar\psi_{h_\pm}&=\bar G^+_{-1/2}\phi_{h_\pm}\ ,\\
    \phi^{\textrm{top}}_{h_\pm}&=G^-_{-1/2}\bar G^+_{-1/2}\phi_{h_\pm} \ .
\end{align}
Naturally we also have the anti-chiral multiplets alongside.
We then explicitly have the following relation of notation:
\begin{align}\label{}
    \phi_{h_\pm}&=T^\pm\mathcal{O}^{1[-]}_{0\pm}\ , & \phi^{\textrm{top}}_{h_\pm}&=-\frac{(2h_\pm-1)^2}{h_\pm}T^\pm\mathcal{O}^{1[+]}_{0\mp}\ ,\\
    \psi_{h_\pm}&=\pm\sqrt{2}\frac{2h_\pm-1}{h_\pm}T^\pm\mathcal{O}^{3/2[+]}_{1/2-}\ , &     \bar\psi_{h_\pm}&=\pm\sqrt{2}\frac{2h_\pm-1}{h_\pm}T^\pm\mathcal{O}^{3/2[-]}_{-1/2+}\ .
\end{align}
Where $T^+$ is the identity and $T^-$ puts a tilde on the operator. While for the anti-chiral multiplets, we have
\begin{align}\label{}
    \tilde\phi_{h_\pm}&=T^\mp\mathcal{O}^{1[-]}_{0\pm}\ , & \tilde\phi^{\textrm{top}}_{h_\pm}&=-\frac{(2h_\pm-1)^2}{h_\pm}T^\mp\mathcal{O}^{1[+]}_{0\mp}\ ,\\
    \tilde\psi_{h_\pm}&=\mp\sqrt{2}\frac{2h_\pm-1}{h_\pm}T^\mp\mathcal{O}^{3/2[+]}_{1/2+}\ , &     \tilde{\bar\psi}_{h_\pm}&=\mp\sqrt{2}\frac{2h_\pm-1}{h_\pm}T^\mp\mathcal{O}^{3/2[-]}_{-1/2-}\ .
\end{align}

We start by considering how the supersymmetry algebra determines the relation between the two-point functions. From the conjugation structure in \eqref{eq:lagrangianstructure}, we see that the possible non-zero two-point functions are
\begin{align}\label{}
    \langle\mathcal{O}^{1[-]}_{0\sigma}\mathcal{\widetilde{O}}^{1[-]}_{0\sigma}\rangle\ , \quad \langle\mathcal{O}^{3/2[+]}_{1/2\sigma}\mathcal{\widetilde{O}}^{3/2[+]}_{1/2-\sigma}\rangle\ , \quad \langle\mathcal{O}^{3/2[-]}_{-1/2\sigma}\mathcal{\widetilde{O}}^{3/2[-]}_{-1/2-\sigma}\rangle\ , \quad \langle\mathcal{O}^{1[+]}_{0\sigma}\mathcal{\widetilde{O}}^{1[+]}_{0\sigma}\rangle\ .
\end{align}
From the CFT point of view, this is just saying that we need to combine a fundamental representation with an anti-fundamental to get the identity representation.

We can now find the relation between these correlators using the supersymmetry Ward identities
\begin{align}\label{}
      \frac{1}{2\pi i}\oint dz\langle \epsilon(z) G^\pm (z) \mathcal{O}\rangle=0\ ,
\end{align}
where $\epsilon(z)$ is maximally linear and the integral encircles all the operators denoted by $\mathcal{O}$. We note that a simple zero can be chosen in $\epsilon(z)$ to avoid an operator having a simple pole OPE with the supercurrents.

With the OPEs in appendix \ref{sec:states} the relations are
\begin{align}\label{eq:relatwopoint}
\begin{split}
    \langle\psi_{h_\pm}(z)\tilde\psi_{h_\pm}(w)\rangle&=-2\del_w\langle\phi_{h_\pm}(z)\tilde\phi_{h_\pm}(w)\rangle\ ,\\
    \langle\bar\psi_{h_\pm}(z)\tilde{\bar\psi}_{h_\pm}(w)\rangle&=-2\bar\del_{\bar w}\langle\phi_{h_\pm}(z)\tilde\phi_{h_\pm}(w)\rangle\ ,\\
    \langle\phi^{\textrm{top}}_{h_\pm}(z)\tilde\phi^{\textrm{top}}_{h_\pm}(w)\rangle&=-4\del_w\delbar_{\bar w}\langle\phi_{h_\pm}(z)\tilde\phi_{h_\pm}(w)\rangle\ ,
\end{split}
\end{align}
or without coordinates
\begin{align}\label{}
    \langle\psi_{h_\pm}(\infty)\tilde\psi_{h_\pm}(0)\rangle&=\langle\bar\psi_{h_\pm}(\infty)\tilde{\bar\psi}_{h_\pm}(0)\rangle=-4h_\pm\langle\phi_{h_\pm}(\infty)\tilde\phi_{h_\pm}(0)\rangle\ ,\\
    \langle\phi^{\textrm{top}}_{h_\pm}(\infty)\tilde\phi^{\textrm{top}}_{h_\pm}(0)\rangle&=-(4h_\pm)^2\langle\phi_{h_\pm}(\infty)\tilde\phi_{h_\pm}(0)\rangle\ .
\end{align}
Note that we could also have done this directly in the bulk theory by relating solutions of the bulk equations of motion, but the CFT method is more familiar to us.
In terms of the bulk terminology this e.g. means
\begin{align}\label{}
    \frac{1}{(2h_+)^4}\langle\mathcal{O}^{1[+]}_{0-}(\infty)\tilde{\mathcal{O}}^{1[+]}_{0-}(0)\rangle=   - \frac{1}{(2(h_+-1/2))^4}\langle\mathcal{O}^{1[-]}_{0+}(\infty)\tilde{\mathcal{O}}^{1[-]}_{0+}(0)\rangle\ .
\end{align}

\subsubsection*{Bosonic projection}

In the bosonic projection of the bulk theory, we only keep operators commuting with $k$, and further project onto an eigenspace of $k$. For the CFT states we keep
\begin{align}\label{}
    P^+:\  \phi_{h_+}\ , \tilde\phi_{h_+}\ ,\ \phi^{\textrm{top}}_{h_-}\ ,\ \tilde\phi^{\textrm{top}}_{h_-}\ ,
\end{align}
for the projection onto $k=+1$ and
\begin{align}\label{}
    P^-:\  \phi_{h_-}\ , \tilde\phi_{h_-}\ ,\ \phi^{\textrm{top}}_{h_+}\ ,\ \tilde\phi^{\textrm{top}}_{h_+}\ ,
\end{align}
for the projection onto $k=-1$. For the symmetry currents we keep $J^{(s)+}$ which in the projection is equal to $\pm J^{(s)-}$. Below we will directly see how the symmetries of the bosonic CFT is embedded into the supersymmetric coset theory.

\subsubsection*{Three-point functions}

We can now easily explain the bulk results for the correlators using right-moving supersymmetry transformations.
The idea used in \cite{AKP} on the bulk side was to get the three-point function by
starting from a two-point function and making a gauge transformation.  In the CFT language this is the Ward identity
\begin{align}\label{}
    &\langle\phi_{h_\pm}(z_1)\tilde\phi_{h_\pm}(z_2)J^{(s)+}(z_3)\rangle=    \frac{1}{2\pi i}\oint_{z_3} dz \frac{1}{z-z_3}\langle\phi_{h_\pm}(z_1)\tilde\phi_{h_\pm}(z_2) J^{(s)+}(z)\rangle  \\
    &=-\frac{1}{2\pi i}\oint_{z_1} dz \frac{1}{z-z_3}\langle J^{(s)+}(z)\phi_{h_\pm}(z_1)\tilde\phi_{h_\pm}(z_2)\rangle-\frac{1}{2\pi i}\oint_{z_2} dz \frac{1}{z-z_3}\langle \phi_{h_\pm}(z_1) J^{(s)+}(z)\tilde\phi_{h_\pm}(z_2)\rangle \ . \nonumber
\end{align}

To get correlators involving fermions from the bosonic three-point functions we do
a supersymmetry transformation using the right-moving versions of the OPEs in appendix \ref{sec:states}
\begin{align}\label{}
  & \langle\bar\psi_{h_\pm}(z_1)\tilde{\bar\psi}_{h_\pm}(z_2)J^{(s)+}(z_3)\rangle=\frac{1}{2\pi i}\oint_{\bar z_1} d\bar z\langle  \bar G^-(\bar z)\phi_{h_\pm}(z_1)\tilde{\bar\psi}_{h_\pm}(z_2)J^{(s)+}(z_3)\rangle \\
  & =-\frac{1}{2\pi i}\oint_{\bar z_2} d\bar z  \langle\phi_{h_\pm}(z_1)\bar G^-(\bar z)\tilde{\bar\psi}_{h_\pm}(z_2)J^{(s)+}(z_3)\rangle=-2\delbar_{\bar z_2}\langle\phi_{h_\pm}(z_1)\tilde{\phi}_{h_\pm}(z_2)J^{(s)+}(z_3)\rangle\ . \nonumber
\end{align}
The point is here that the right moving supercurrent does not have an OPE with the left-moving higher spin current. Now, knowing that
\begin{align}\label{}
    \langle\phi_{h_\pm}(z_1)\tilde{\phi}_{h_\pm}(z_2)J^{(s)+}(z_3)\rangle=A_{h_\pm}(z_1,z_2,z_3)\langle\phi_{h_\pm}(z_1)\tilde{\phi}_{h_\pm}(z_2)\rangle \ ,
\end{align}
we directly get from the comparison of two-point functions in eq. \eqref{eq:relatwopoint}
\begin{align}\label{}
   \langle\bar\psi_{h_\pm}(z_1)\tilde{\bar\psi}_{h_\pm}(z_2)J^{(s)+}(z_3)\rangle=A_{h_\pm}(z_1,z_2,z_3)\langle\bar\psi_{h_\pm}(z_1)\tilde{\bar\psi}_{h_\pm}(z_2)\rangle \ .
\end{align}
This is exactly the result obtained on the bulk side, i.e. that correlators with fermions have the same pre-factor as the bosonic correlators.
We also need to show this for the fermionic states $\psi_{h_\pm}$. Relating to correlators with $\phi_{h_\pm}$ would not give such a simple relation since the left-moving supercurrent would also have an OPE with the higher spin current. However in the comparison of the bosonic result \eqref{bosonresult} and the fermionic result \eqref{sugraresult} we see that we exactly should relate to the top components. We then have in the same way
\begin{align}\label{}
    \langle\phi^{\textrm{top}}_{h_\pm}(z_1)\tilde{\phi}^{\textrm{top}}_{h_\pm}(z_2)J^{(s)+}(z_3)\rangle=2\delbar_{\bar z_2}   \langle\psi_{h_\pm}(z_1)\tilde{\psi}_{h_\pm}(z_2)J^{(s)+}(z_3)\rangle\ .
\end{align}
Given that
\begin{align}\label{}
   \langle\psi_{h_\pm}(z_1)\tilde{\psi}_{h_\pm}(z_2)J^{(s)+}(z_3)\rangle=B_{h_\pm}(z_1,z_2,z_3)\langle\psi_{h_\pm}(z_1)\tilde{\psi}_{h_\pm}(z_2)\rangle
\end{align}
we thus again conclude that the coefficients have to be the same for the bosonic correlators i.e.
\begin{align}\label{}
\langle\phi^{\textrm{top}}_{h_\pm}(z_1)\tilde{\phi}^{\textrm{top}}_{h_\pm}(z_2)J^{(s)+}(z_3)\rangle= B_{h_\pm}(z_1,z_2,z_3)   \langle\phi^{\textrm{top}}_{h_\pm}(z_1)\tilde{\phi}^{\textrm{top}}_{h_\pm}(z_2)\rangle\ .
\end{align}

Let us finally show that we can also get the correlators with a fermionic current via supersymmetry. Let us for simplicity consider the correlator with the boson $\phi_{h_+}$, the fermion $\tilde{\psi}_{h_+}$ and thus the current $W^{s-}$. We find via the Ward identity
\begin{multline}\label{}
    \langle\phi_{h_+}(z_1)\tilde{\psi}_{h_+}(z_2)W^{s-}(z_3)\rangle =\langle\psi_{h_+}(z_1)\tilde{\psi}_{h_+}(z_2)W^{s0}(z_3)\rangle \\+ 2\del_{z_2}\langle\phi_{h_+}(z_1)\tilde{\phi}_{h_+}(z_2)W^{s0}(z_3)\rangle \ .
\end{multline}
Using the Ward identity with a linear parameter that is zero in $z_3$, we can relate the correlator with the fermions to that with bosons. We then get
\begin{align}
    \langle\phi_{h_+}(z_1)\tilde{\psi}_{h_+}(z_2)W^{s-}(z_3)\rangle&=2\frac1{z_{13}}(z_{12}\del_{z_2}-2 h_+)\langle\phi_{h_+}(z_1)\tilde{\phi}_{h_+}(z_2)W^{s0}(z_3)\rangle \nonumber \\
    &=-\frac{2s}{z_{23}}\langle\phi_{h_+}(z_1)\tilde{\phi}_{h_+}(z_2)W^{s0}(z_3)\rangle \ ,\label{3ptfc}
\end{align}
where in the last equation we have used that the coordinate dependence of the three-point function is fixed.

\subsection{Recursion relations}
\label{sec:recursion}

We can now in principle calculate all the correlators related by supersymmetry, i.e. within the supermultiplets. However, on the bulk side we know that in correlators the value of $k$ is fixed by the matter, $k=\pm 1$. This means that for our correlators, we have a relation between the two bosonic spin-s generators $J^{(s)-}=\pm J^{(s)+}$. In this section we will assume this to be true in the CFT theory also. We can then easily obtain a relation between the correlators with a spin $s$ and a spin $s+1$ current. Indeed, using \eqref{eq:susytranswalg} we get
\begin{align}\label{}
    0&=\frac{1}{2\pi i}\oint_{z_3} dz \frac{z-z_2}{z_3-z_2}\langle G^{+}(z)\phi_{h_\pm}(z_1)\tilde{\phi}_{h_\pm}(z_2)W^{s-}(z_3) \rangle \nonumber \\
    &=\langle \phi_{h_\pm}(z_1)\tilde{\phi}_{h_\pm}(z_2)\big(\frac{2s}{z_3-z_2}W^{s0}(z_3)+2W^{s1}(z_3)+\del_{z_3}W^{s0}(z_3)\big) \rangle\ .
\end{align}
Using \eqref{eq:cftcurrentsinbulknota} and that $k=\pm1$, we then get the recursion relation
\begin{multline}\label{}
    \langle \phi_{h_\pm}(z_1)\tilde{\phi}_{h_\pm}(z_2)J^{(s+1)+}(z_3) \rangle \\ = -\frac12\left(\frac\nu{4(s-1/2)}\pm\frac12\right)\left(\frac{2s}{z_3-z_2}+\del_{z_3}\right)\langle \phi_{h_\pm}(z_1)\tilde{\phi}_{h_\pm}(z_2)J^{(s)+}(z_3) \rangle \ .
\end{multline}
For the spin one case we can use that $W^{10}=U$ and $W^{10}=(\nu J^{(1)+}+J^{(1)-})/2$ to calculate
\begin{align}\label{}
    \langle \phi_{h_\pm}(z_1)\tilde{\phi}_{h_\pm}(z_2)J^{(1)+}(z_3) \rangle=\pm\frac{z_{12}}{z_{13}z_{23}}\langle \phi_{h_\pm}(z_1)\tilde{\phi}_{h_\pm}(z_2) \rangle \ .
\end{align}
This is the same result as obtained in \cite{AKP} up to the factor of $2\pi$ which comes from bulk-boundary coupling. Performing the induction step we now finally obtain
\begin{align}\label{}
    \langle \phi_{h_\pm}(z_1)\tilde{\phi}_{h_\pm}(z_2)J^{(s)+}(z_3) \rangle=-(\mp)^{s}\frac{\Gamma(s)^2\Gamma(s-\lambda_\pm)}{\Gamma(2s-1)\Gamma(1-\lambda_\pm)}\left(\frac{z_{12}}{z_{13}z_{23}}\right)^s \langle \phi_{h_\pm}(z_1)\tilde{\phi}_{h_\pm}(z_2) \rangle \ ,
\end{align}
which is the result conjectured in \cite{AKP} (up to the $2\pi$ factor).

We have thus seen that considering the untruncated supersymmetric theory provides us with much stronger symmetry than the bosonic truncation. In particular, the supersymmetry algebra along with the knowledge of how the multiplication with $k$ works on the bulk side, gives us the result in a very simple way. Note that on the bulk side the multiplication with $k$ can be obtained in the Lie superalgebra as follows: For the fermionic operators, simply consider the commutator with $k$, for the bosonic operators consider the commutators with $V^{(2)-}_m$. Indeed, it was shown in appendix \ref{ref:appsupertrace} that the supertrace is determined by all generators with spin 2 and less. This leads us to suspect that the currents of spin 1, 3/2 and 2 generate the whole super ${\cal W}[\lambda]$ algebra as we will show in the following.

\subsection{Symmetries of the coset CFT}

\label{symmetry}

In this subsection, we give an explicit realization of generators of the symmetry algebra.
Consider the affine Lie algebra $\widehat{\text{su}}(N+1)_k$.
It decomposes as
\begin{equation}
\widehat{\text{su}}(N+1)_k \ = \ \widehat{\text{su}}(N)_k \oplus \widehat{\text{u}}(1) \oplus V_N \oplus V'_N \ ,
\end{equation}
where $V_N$ denotes the $N$-dimensional fundamental representation of $\widehat{\text{su}}(N)_k$  and $V'_N$ is its conjugate.
Denote the corresponding fields by $(J^a,\tilde J,B^\pm_i)$.
We view the $2N$ real fermions as $N$ complex ones, then
the (linear) fermions themselves decompose into the fundamental and anti-fundamental
representation of $\widehat{\text{su}}(N)_1$, while the bilinears in the fermions are $\widehat{\text{su}}(N)_1 \oplus \widehat{\text{u}}(1)$.
Denote the fields by $(j^a,\tilde j,\psi^\pm_i)$. Then the coset algebra is the subalgebra of the symmetry algebra of the parent CFT that
commutes with the symmetry algebra of the theory we quotient by.
In our case this means we are looking for fields that commute with $\widehat{\text{su}}(N)_{k+1}\oplus \widehat{\text u}(1)$, i.e. with the currents
\begin{equation}
K^a\ = \ J^a+j^a \ ,\qquad \tilde K\ = \ \tilde J+\tilde j\ .
\end{equation}
We find the following elements that, as we will explain in the next subsection, already generate the complete symmetry algebra under iterated operator products;
\begin{equation}\label{generators}
\begin{split}
U &= \frac{1}{N+k+1}(\tilde J-\frac{k}{N+1}\tilde j) \ ,\qquad
W= T_{\widehat{\text{su}}(N)_k}+T_{\widehat{\text{su}}(N)_1}-T_{\widehat{\text{su}}(N)_{k+1}} \ ,\\
G^\pm&= \sum_i B^\pm_i\psi^\mp_i \ , \qquad T = T_{\widehat{\text{su}}(N+1)_k}+T_{\text{fermion}}- T_{\widehat{\text{su}}(N)_{k+1}}-T_{\tilde{K}} \ .
\end{split}
\end{equation}
The first one is the obvious U(1)-current with normalization from \eqref{eq:superconfalgope} and calculated using that \mbox{$\tilde J(z) \tilde J(w)\sim N(N+1)k/(z-w)$} and $\tilde j(z)\tilde j(w)\sim N(N+1)^2/(z-w)$. The following two fermionic dimension $3/2$ fields are the invariants of the tensor product of the fundamental representation
with its conjugate and since $B^\pm$ and $\psi^\pm$ commute, this implies them being in the commutant.
Finally, the dimension $2$ field $T$ is the Virasoro field of the super coset, while the dimension $2$ field $W$ is the Virasoro field of the {\emph{bosonic}} coset of the theory, i.e. of the coset
\begin{align}\label{scoset}
 \frac{\widehat{\text{su}}(N)_k \oplus \widehat{\text{su}}(N)_1 }{\widehat{\text{su}}(N)_{k+1} }\  .
\end{align}
Actually, any field of the symmetry algebra of the above {\emph{bosonic}} coset is also a field of the symmetry algebra of the supersymmetric coset.
The reason is, that $\widehat{\text{su}}(N)_k \oplus \widehat{\text{su}}(N)_1$ is a subalgebra of $\widehat{\text{su}}(N+1)_k\oplus \text{fermions}$
that commutes with the $\widehat{\text{u}}(1)$ of the nominator. Hence, the symmetry algebra of the supercoset restricted to this subalgebra is the symmetry algebra of the  {\emph{bosonic}} coset. The latter has the bosonic ${\cal W}_N$ algebra as symmetry algebra, that is for each spin $s=2,...,N$ one generator which we denote $W^s_b$.
$W$ is not a primary, since the operator product with $T$ is
\begin{equation}
 T(z)W(w)\sim \frac{c_b/2}{(z-w)^4}+\frac{2W(w)}{(z-w)^2}+\frac{\partial W(w)}{(z-w)}
\end{equation}
where $c_b$ the central charge of the bosonic coset \eqref{scoset}.
Using this OPE and \eqref{eq:superconfalgope} we see that the field
\begin{align}\label{}
    W^{20}&=W+\frac{c_b}{1-c}\bigl( T-\frac{3}{2c} :UU:\bigr)
\end{align}
is primary and has vanishing operator product with $U$. It is thus the field that is the bottom component of the $\mathcal N=2$ supermultiplet, however now even in the finite $N$ case.
In the large $N$ limit we have $c_b\sim N(1-\lambda^2)$ and $c\sim 3(1-\lambda)N$ and hence
\begin{align}\label{eq:w20construct}
    W^{20}&=W-\frac{1+\lambda}{3}\bigl( T-\frac{3}{2c} :UU:\bigr) \ .
\end{align}
This is exactly what we expect from the bulk side, up to the $:UU:$ which is zero for finite $U$ charges. The point is that the bosonic hs$[\lambda]$ subalgebra is generated by $P_+ V^{(s)+}$ with dual currents $(J^{(s)+}+J^{(s)-})/2$, whereas $T$ is $J^{(2)+}$ and $W^{20}$ by \eqref{eq:cftcurrentsinbulknota} is
\begin{align}\label{}
    W^{20}=\frac{(1-2\lambda)J^{(2)+}+3J^{(2)-}}{6} \ ,
\end{align}
which exactly solves to \eqref{eq:w20construct}.

Thus to provide a check of the bulk fact that $k=\pm 1$ in the correlators, which we used successfully in last section, we need to show that $W$ on our matter states act as $T$ or zero.
We will thus give an explicit mapping of the matter states to the bosonic theory.
First to leading level, the identity representation $s=0$ of $\widehat{\text{so}}(2N)_1$ transforms in the trivial representation of
$\widehat{\text{su}}(N)_1$, while the vector
representation, $s=2$, transforms in the fundamental plus anti-fundamental representation of $\widehat{\text{su}}(N)_1$.
Since in the nominator, the $\widehat{\text{su}}(N)_{k+1}$ are the same, primaries also transform in the same representation. Further, the (anti-)fundamental representation of
$\widehat{\text{su}}(N+1)_k$ decomposes into the (anti-)fundamental and the trivial representation of $\widehat{\text{su}}(N)_k$ and the trivial module of course remains trivial. We then obtain
\begin{align}\label{}
(\text f,2;0)  &\longrightarrow \, (\text f,\bar{\text{f}};0)_b\ , &  (0,2;\text f) & \longrightarrow \, (0,\text f;\text f)_b, \\
 (\text f,0;0) &\longrightarrow \, (0,0;0)_b\ , &  ( 0,0;\text{f}) &\longrightarrow \, (\text{f},0;\text{f})_b\ ,
\end{align}
where for the last state we have used that it appears on the second level.
In fact, this was already used in \cite{CHR} when we calculated its conformal weight. These identifications were also obtained in that paper when we expanded the partition function to low orders. The two upper states are the $k=1$ states and we indeed see that these have the same conformal weights for the full and the bosonic Virasoro tensor. The two lower states have $k=-1$ and they nicely have conformal weight zero in the large $N$ limit.

\subsection{Generating fields of the symmetry algebra }

\label{generating}

We claimed that the fields of \eqref{generators} already generate all other fields of the symmetry algebra under iterated operator products.
We know that the bosonic subalgebra is generated by the fields of spin $1,2,3$, see e.g. Lemma 4.1 of \cite{Linshaw}. We also know that the
bosonic and fermionic generators combine into multiplets of the $\mathcal N=2$ superconformal algebra. Hence, if $U, G^\pm, T, W$ generate
the spin three fields under OPE, then they already generate the complete algebra.
Let us take the limit $k\rightarrow \infty$. In that limit the invariant fields can be described as the U($N$) invariants of $N$ pairs of fermions
$b_i,c_i$ and $N$ pairs of bosons $\partial X_i,\partial Y_i$ with operator products
$$ b_i(z)c_j(w) \sim \frac{\delta_{i,j}}{(z-w)} \ , \qquad  \partial X_i(z)\partial Y_j(w) \sim \frac{\delta_{i,j}}{(z-w)^2} \ .$$
Here $b$ and $Y$ carry the fundamental representation of u($N$), and $c$ and $X$ the conjugate representation.
The invariants of spin $1,3/2,2$ are
$$ :b_ic_i: , \qquad  :b_i\partial X_i:, \qquad :c_i\partial Y_i, \qquad  :b_i\partial c_i:, \qquad :c_i\partial b_i:, \qquad :\partial X_i\partial Y_i: .$$
We compute the following contributions to the operator product
\begin{align}
 :c_i\partial b_i:(z):c_i\partial Y_i:(w) &\sim \cdots + \frac{:\partial c_i\partial Y_i:(w)}{(z-w)} \ , \\
:b_i\partial X_i:(z):\partial c_i\partial Y_i:(w) &\sim  \cdots +\frac{-:\partial^2X_i\partial Y_1:(w)+:\partial b_i\partial c_i:(w)}{(z-w)} \ , \nonumber  \\
:\partial X_i\partial Y_i:(z):\partial^2X_i\partial Y_1-\partial b_i\partial c_i:(w)&\sim \cdots +\frac{3:\partial^2X_i\partial Y_1:(w)}{(z-w)^2}+ \cdots \ ,  \nonumber
\end{align}
where the dots denote contributions from other poles. These operator products show that the spin three fields $:\partial^2X_i\partial Y_1:$, $:\partial b_i\partial c_i:$
appear. We have thus established that in the large $k$ limit the symmetry algebra is generated by the spin 1, 3/2 and 2 fields. The same statement is true for generic finite level $k$, as
one can continuously deform the operator product algebra, see \cite{deBoer:1993gd}.

\section{Conclusion and outlook}
\label{conclusion}

In \cite{CHR} we have proposed that the higher spin ${\cal N}=2$ supergravity on AdS$_3$
constructed in \cite{PV1} is dual to the 't Hooft limit of the $\mathbb{C}\text{P}^N$ Kazama-Suzuki
model \eqref{sminimal}
\begin{align}
  \frac{\widehat{\text{su}}(N+1)_k \oplus \widehat{\text{so}}(2N)_1 }
{\widehat{\text{su}}(N)_{k+1}  \oplus \widehat{\text{u}}(1)_{N(N+1)(k+N+1)}} \nonumber \ .
\end{align}
This conjecture has been supported by the analysis of symmetry and spectrum.
In this paper, we have examined correlation functions to add more evidence.
Concretely, we have computed boundary
three point functions with two fermionic operators and one bosonic higher spin current from the
dual supergravity theory by applying a method in \cite{AKP} used for the bosonic duality.
The results are summarized in eq. \eqref{sugraresult} and shown
to be a result of supersymmetry in the CFT analysis.

It is useful to observe a relation between the two bosonic currents of spin $s$
when acting on the dual matter states, which is evident on the bulk side.
Using the relation and the supersymmetry, we obtain
a recursion relation between correlators of currents with spin $s$ and $s+1$. This recursion relation reproduces the
previously conjectured result of \cite{AKP}. Further, we constructed the $\mathcal{N}=2$ supersymmetry algebra explicitly in the super coset
theory together with the second current of spin two via an identification of how the bosonic $\mathcal{W}[\lambda]$ algebra is obtained as a sub-algebra. We also showed that these spin two currents have the expected relation on the matter states. Finally, we have proven that the currents of spin 1, $3/2$ and 2 generate the whole super $\mathcal{W}[\lambda]$ algebra. We thus expect that all higher spin currents also have the correct relations on the matter states, but have postponed this analysis to future studies.

In \cite{CHR3} we have also proposed the ${\cal N}=1$ version of the duality,
and the analysis in this paper can easily be applied to that case.
This is because the gravity theory is obtained by the ${\cal N}=1$ truncation of
the ${\cal N}=2$ supergravity \cite{PV1}, while the ${\cal N}=1$ supersymmetry
of the dual CFT  can be treated as a sub-algebra of the ${\cal N}=2$ supersymmetry.

There are several other open problems worth studying. On the CFT side we have used supersymmetry to calculate correlation functions involving a
fermionic gauge field $J^{(s+1/2)}$ like
\begin{align}
\left \langle  {\mathcal O}_B^{(h, h)}(z_1)  \widetilde {\mathcal O}_F^{(h \pm 1/2,h)}(z_2 ) J^{(s+1/2)} (z_3) \right \rangle
\end{align}
with $s \in \mathbb{Z}$,
see \eqref{3ptfc}.
This result should be obtained by a direct computation from the supergravity theory. The necessary structure constants of the higher spin algebra have already been calculated in appendix \ref{ref:appsupertrace}.

In this paper, we have focused on the 't Hooft limit of the $\mathbb{C}\text{P}^N$ model,
but it is important to study the $1/N$ corrections.
Applying the duality, we can examine the quantum effects of supergravity from the $1/N$ expansions of
the dual CFT, and these effects could be more tractable in our supersymmetric setup.
For instance, we can compute three point function with one higher spin current where $k,N$ are kept finite, in principle.
Other correlation functions would be important as well.
In \cite{Papadodimas:2011pf,Chang:2011vk} four point functions of scalar operators
are investigated, and it was argued that some extra states would appear if $1/N$ effects are included.
We would expect similar things to happen in our case.
Finally, by introducing supersymmetry we may be able to see the relation to superstring
theory as discussed in \cite{Henneaux:2012ny}, since
higher spin supergravity is believed to be related to the tensionless limit of
superstring theory.

\subsection*{Acknowledgements}

We are grateful to H.~Moradi and K.~Zoubos for sharing their draft \cite{MZ}.
The work of YH was supported in part by Grant-in-Aid for Young Scientists (B) from JSPS, and the work of PBR is funded by DFG grant no. ZI 513/2-1.

\appendix

\section{Higher spin algebras}
\label{hsas}

In this appendix, we review some useful facts on the higher spin algebras hs$[\lambda]$ and
shs$[\lambda]$.

\subsection{Structure constants of hs$[\lambda]$}

The higher spin algebra hs$[\lambda]$ are generated by $V_m^s$ with $s=2,3,\ldots$
and $|m| = 0 , 1 , \ldots, s-1$. The commutation relations among the generators are
\begin{align}
 [ V_m^s , V_n^t] = \sum_{u=2,4,\cdots}^{s+ t - |s - t| - 1} g_{u}^{st} (m,n;\lambda) V_{m+n}^{s+t-u} ~,
\end{align}
and the structure constants are given as \cite{PRS}
\begin{align}
 g_{u}^{st} (m,n;\lambda) = \frac{q^{u-2}}{2 (u-1)!} \phi_u^{st} ( \lambda ) N_u^{st} (m,n)
 ~ .\label{gust}
\end{align}
Here we have defined
\begin{align} \nonumber
& N_u^{st} (m,n) = \sum_{k=0}^{u-1} (-1)^k \begin{pmatrix} u-1 \\ k \end{pmatrix}
 [ s-1+m]_{u-1+k } [s - 1 - m]_k [t - 1 + n]_k [t - 1 - n]_{u-1-k} ~, \\
& \phi_u^{st} ( \lambda)
 = {}_4 F_3 \left[ \begin{matrix} \frac{1}{2} + \lambda , \frac{1}{2} - \lambda  ,
     \frac{2 - u}{2}  ,   \frac{1 - u}{2}  \\
  \frac{3}{2} - s , \frac{3}{2} - t , \frac{1}{2} + s + t - u \end{matrix} \right| 1 \biggr]
  \label{Nphi}
\end{align}
with $[a]_n = \Gamma(a+1)/\Gamma(a+1-n)$. We set the normalization constant as $q=1/4$.

\subsection{Structure constants of shs$[\lambda]$}

We can generalize the higher spin algebra hs$[\lambda]$ by incorporating ${\cal N}=2$
supersymmetry \cite{Bergshoeff:1990cz,Bergshoeff:1991dz}. The algebra may be called as shs$[\lambda]$ as in \cite{CHR}, and
it is generated by
\begin{align}
V_n^{(s)+} ~ (s=2,3,\cdots) \ , \quad V_n^{(s)-} ~ ( s=1,2,\cdots)  \ , \quad
F^{(s)\pm}_r\equiv V^{(s+1/2)\pm}_{r} ~ (s = 1, 2,\cdots)
\end{align}
with $|n| = 0 ,1 ,\ldots , s-1, |r| = 1/2 , 3/2 , \ldots , s - 1/2$.
The generators
$V^{(2)+}_0$, $V^{(2)+}_{\pm 1}$, $F^{(1)+}_{\pm1/2}$
form  a basis of osp$(1|2)$ subalgebra as
\begin{align}\label{eq:N1algebra}
& [V^{(2)+}_m , V^{(2)+}_n] = (m-n) V_{m+n}^{(2)+}~,  \qquad
 [V^{(2)+}_m , F^{(1)+}_r] = (\tfrac{1}{2} m - r) F^{(1)+}_{m+r} ~, \nonumber \\
& \{ F^{(1)+}_r, F^{(1)+}_s\} = 2 V^{(2)+}_{r+s} ~.
\end{align}
Among the other generators, (anti-)commutation relations are
\begin{align}\label{eq:commuosp12}
&[V_m^{(2)+} , V_n^{(s)\pm}] = (- n + m (s-1) ) V_{m+n}^{(s)\pm} ~, \qquad
[V_m^{(2)+} , F_r^{(s)\pm}] = (- r + m (s-\tfrac{1}{2}) )F_{m+n}^{(s)\pm} ~, \nonumber \\
&[F_{1/2}^{(1)+} , V_m^{(s)+}] = - \tfrac12 (m -s +1) F_{m+1/2}^{(s-1)+} ~, \qquad
[F_{1/2}^{(1)+} , V_m^{(s)-}] = - 2 F_{m+1/2}^{(s)-} ~, \\
&\{ F^{(1)+}_{1/2} , F_r^{(s-1)+}\} = 2 V_{r+1/2}^{(s)+} ~, \qquad
\{ F^{(1)+}_{1/2} , F^{(s)-}_r \} = \tfrac12 (r-s +\tfrac{1}{2}) V_{r+1/2}^{(s)-}  ~.  \nonumber
\end{align}
Here the labels take $n,m \in \mathbb{Z}$ and $r \in \mathbb{Z} + 1/2$
satisfying $|n| , |m| \leq s - 1$ and $|r| \leq s - 1/2  $.
We can show that $k+\nu , F^{(1)\pm}_{\pm 1/2} , V^{(2)+}_0 , V^{(2)+}_{\pm 1} $
generate osp$(2|2)$ subalgebra.
The other commutation relations can be found in \cite{Bergshoeff:1991dz}.

\section{Star product approach to higher spin algebras}
\label{structure}

In this appendix we introduce the star product on the shs[$\lambda$] and use it for some explicit calculations.

\subsection{The star product}

The superalgebra shs[$\lambda$] is generated by $\tilde y_\alpha,k$ with
\begin{align}
 [\tilde y_\alpha , \tilde y_\beta] = 2 i \epsilon_{\alpha \beta} (1 + \nu k) \, , \qquad
 \{ k , y_\alpha \} = 0
 \label{yycom}
\end{align}
and $\epsilon_{12} = - \epsilon_{21} = 1$.
We express the generators as
\begin{align}
 V^{(s)+}_m = \left( \frac{-i}{4}\right)^{s-1} S^s_m ~, \qquad
 V^{(s)-}_m &=  \left( \frac{-i}{4}\right)^{s-1} k S^s_m ~,
\end{align}
where $S_m^s$ are symmetric products of $\tilde y_\alpha$.
Denoting the numbers of $\tilde y_{1,2}$ as $N_{1,2}$,
the indices are
\begin{align}
 N_1 + N_2 = 2 s - 2 \, , \qquad N_1 - N_2 = 2 m ~.
\end{align}
For a short while, we ignore the effect of $k$ and set $V^s_m =  V^{(s)+}_m$.
The star products among $V_m^s$ can be expressed as \eqref{lsp}
\begin{align}
 V_m^s * V_n^t = \frac12 \sum_{u=1,2,\cdots}^{s+t-|s-t| -1}
 g^{st}_u (m,n;\lambda_k) V^{s+t-u}_{m+n}
\end{align}
with $\lambda_k=(1-\nu k)/2$, i.e. $P_\pm\lambda_k=\lambda_\pm$.
The expression is quite useful for the bosonic subsector with $s,t,m,n \in \mathbb{Z}$,
since the closed form of structure constant is conjectured to be given in \eqref{gust}.
For the case involving also half integer $s,t,m,n$, we have to compute
the coefficients $g^{st}_u (m,n;\lambda_k)$
directly by applying the commutation relation \eqref{yycom} or deduce them from
bosonic ones.

\subsection{Some explicit calculations for $V^{3/2}_m$ and $V^{2}_m$}

In order to derive the field equations for matter fields in the AdS background, we need
to compute the star products between $V^{3/2}_{\pm 1/2} , V^{2}_{0,\pm 1}$
and generic $V^s_m$.
Since the detailed analysis have been done in appendix
C of \cite{CY}, the task now is only to change the basis of the symmetric products from
$y_{( \alpha_1} \cdots y_{ \alpha_{n})}$ into $S^s_m$. For the computation
with the multiplication of $V^{3/2}_{\pm 1/2}$ (or one $y_\alpha$), we may utilize eq. (C.12) of the
paper. By changing the basis we obtain
\begin{align}
 &V^{\frac{3}{2}}_{+\frac12} * V^s_m = V^{s + \frac12}_{m+\frac12} -  a (2s - 2 , \nu k) \frac{m-s+1}{8 (s-1)}
  V^{s-\frac12}_{m+\frac12} ~, \\
 &V^{\frac{3}{2}}_{-\frac12}* V^s_m = V^{s + \frac12}_{m-\frac12} -  a (2s-2 , \nu k) \frac{m+s-1}{8(s-1)}
  V^{s-\frac12}_{m-\frac12}
\end{align}
with
 \begin{align}
  a (n,\nu k ) &= 2 \sum_{i=1}^n \frac{1}{(n+1)} (n-i+1) (1+(-)^{i-1} \nu k) \\
 & = \left \{
  \begin{array}{ll}
   n + \frac{n}{n+1} \nu k & \text{for} ~ n \in 2 \mathbb{Z} ~ , \\
   n + \nu k & \text{for} ~ n \in 2 \mathbb{Z} + 1 ~ .
  \end{array}
    \right. \nonumber
 \end{align}
In the same way we have
\begin{align}\label{eq:V32right}
 &V^s_m * V^{\frac{3}{2}}_{+\frac12} = V^{s + \frac12}_{m+\frac12} - b (2s - 2 , \nu k) \frac{m-s+1}{8(s-1)}
  V^{s-\frac12}_{m+\frac12} ~, \\
 &V^s_m  * V^{\frac{3}{2}}_{-\frac12} = V^{s + \frac12}_{m-\frac12} - b (2s-2 , \nu k) \frac{m+s-1}{8(s-1)}
  V^{s-\frac12}_{m-\frac12}
\end{align}
with
 \begin{align}
  b (n,\nu k ) &= 2 \sum_{i=1}^n \frac{1}{(n+1)} (-i) (1+(-)^{i-1} \nu k) \\
 & = \left \{
  \begin{array}{ll}
  - n + \frac{n}{n+1} \nu k & \text{for} ~ n \in 2 \mathbb{Z} ~ , \\
  - n - \nu k & \text{for} ~ n \in 2 \mathbb{Z} + 1 ~ .
  \end{array}
    \right. \nonumber
 \end{align}

Applying $V^{3/2}_{\pm 1/2}$ (or $y_\alpha$) once again, we obtain the equations similar to (C.15) and
(C.19) of \cite{CY} and from them we can read off the coefficients
$g_u^{st}(m,n;\lambda_k)$  for $s=2$ or $t=2$.
For the bosonic case with $s,t \in \mathbb{Z}$ we can reproduce the
formula in \eqref{gust}. For  $s \in \mathbb{Z}+1/2$, relevant formula are
\begin{align}
 &g^{2s}_2 (0,m;\lambda_k) = - m (1 - \tfrac{1-2\lambda_k}{4s(s-1)}) ~ , \qquad
 g^{s2}_2 (m,0;\lambda_k) =  m (1 + \tfrac{1-2\lambda_k}{4s(s-1)}) ~, \\
 &g^{2s}_2 (1,m;\lambda_k) =  ( s - 1 - m )(1 - \tfrac{1-2\lambda_k}{4s(s-1)}) ~ , \qquad
 g^{s2}_2 (m,1;\lambda_k) = - (s - 1 - m) (1 + \tfrac{1-2\lambda_k}{4s(s-1)}) ~, \nonumber \\
 &g^{2s}_2 (-1,m;\lambda_k) = - (s - 1 + m ) (1 - \tfrac{1-2\lambda_k}{4s(s-1)}) ~ , \qquad
 g^{s2}_2 (m,-1;\lambda_k) =  (s - 1 + m) (1 + \tfrac{1-2\lambda_k}{4s(s-1)}) \nonumber
\end{align}
and
\begin{align}
 &g^{2s}_3 (0,m;\lambda_k) = - \tfrac{1}{32 (s-1)^2}
  (s-1+m)(s-1-m) (2s-1-2 \lambda_k) (2s - 3 + 2 \lambda_k) ~, \\
& g^{2s}_3 (1,m;\lambda_k) = \tfrac{1}{32 (s-1)^2}
  (s-1-m)(s-2-m) (2s-1-2 \lambda_k) (2s - 3 + 2 \lambda_k) ~, \nonumber \\
& g^{2s}_3 (-1,m;\lambda_k) = \tfrac{1}{32 (s-1)^2}
  (s-1+m)(s-2+m) (2s-1-2 \lambda_k) (2s - 3 + 2 \lambda_k) ~. \nonumber
\end{align}
 We can also show that
\begin{align}
 g^{2s}_3 (n,m;\lambda_k)  = g^{s2}_3 (m,n;\lambda_k)
\end{align}
even for $s \in \mathbb{Z}+1/2$.

\subsection{Automorphisms and anti-automorphisms of the higher spin algebra}\label{sec:auto}

As already found\footnote{In comparison with \cite{Bergshoeff:1991dz} we use that we have an isomorphism relating shs[$\lambda$] and shs[$1-\lambda$] via $k\mapsto-k$} in \cite{Bergshoeff:1991dz} we have a $\mathbb{Z}_4$ anti-automorphism of the supersymmetric higher spin algebra which exchanges order and takes
\begin{align}\label{}
    \sigma(\tilde y_\alpha)= i \tilde y_\alpha\ .
\end{align}
The action on the generators are then:
\begin{align}\label{eq:anti-auto}
    \sigma(V_m^{(s)\pm})=(\pm)^{2s}(-1)^{s-1}V_m^{(s)\pm}\ .
\end{align}
In order to see the $k$-dependence more explicitly,
we use a bit different notation for the coefficients of the star-algebra as
\begin{align}
 V_m^{(s)+} * V_n^{(t)+} = \frac12 \sum_{u=1,2,\cdots}^{s+t-|s-t| -1}
 g^{st}_u (m,n;\lambda,k) V^{(s+t-u)+}_{m+n}
\end{align}
with $\lambda = \lambda_+ = (1 - \nu)/2$.
From these coefficients star products involving $V_m^{(s)-}$ are trivial to obtain.
Due to the anti-automorphism these coefficients fulfill
\begin{align}\label{eq:conjcoeffi}
    g^{st}_u (m,n;\lambda,k)=(-1)^{1+u}g^{ts}_u (n,m;\lambda,(-1)^{2(t+s)}k)\ .
\end{align}
To get the action on the fields, we demand that the equations of motion \eqref{feforc} are kept invariant. We thus demand that the order of fields gets exchanged, that $\eta$ exchanges $C$ and $\tilde C$, and exchanges signs on $A$ and $\bar A$. On the fields we then get
\begin{align}\label{eq:antiautoonCA}
 \eta(C^s_{m,\sigma})=(-1)^{-s+1}\tilde C^s_{m,(-1)^{2s}\sigma}\ ,\qquad \eta(A^s_{m,\sigma})=(-1)^{-s} A^s_{m,(-1)^{2s}\sigma}\ ,
\end{align}
where $A=\sum_{\sigma,s} \sum_{|m| \leq s-1} P_\sigma A^s_{m,\sigma} V^s_m$. Note that this is indeed fulfilled by the AdS$_3$ solution without any changes of coordinates.

We note that the superconformal algebra \eqref{eq:superconfalg} has the same anti-automorphism for its global subalgebra:
\begin{align}\label{}
\begin{split}
    U_0&\mapsto U_{0}\ ,\\
    L_m&\mapsto -L_{m} ,\qquad m=-1,0,1\ ,\\
    G^\pm_{\pm1/2}&\mapsto i G^\mp_{\pm1/2}\ .
\end{split}
\end{align}
Whereas the isomorphism $k\mapsto-k$ and $\lambda\mapsto -\lambda$ descends from the affine automorphism taking $U\mapsto-U$ and $G^\pm\mapsto G^{\mp}$.

We can also realize a $\mathbb{Z}_2$ anti-automorphism which changes order and maps $(\tilde y_1)^{t}=\tilde y_2$, i.e. on generators $(V_m^{(s)+})^{t}= V_{-m}^{(s)+}$. Looking at what happens to the sl$(2|1)$ sub-algebra, we see that this is simply transposition on the finite matrices recovered for $\lambda\in\mathbb{Z}$, and this is the reason that we denote it with transpose. On the CFT side it extends to the standard conjugation on the whole affine algebra taking $L^\dagger_m=L_{-m}$, $(G^{\pm}_m)^\dagger=G^\mp_{-m}$ and $U^\dagger_m=U_{-m}$.

Finally, we can also make a $\mathbb{Z}_4$ automorphism by combining the two anti-automorphisms. Up to a conjugation, we can do this by taking $\tilde y_1\mapsto-\tilde y_2$ and $\tilde y_2\mapsto\tilde y_1$, and $\psi_i\mapsto-\psi_i$. This maps $V_m^{(s)\pm}\mapsto (-1)^{m+s-1} V_{-m}^{(s)\pm}$. We then demand that this maps $C\mapsto\tilde C$ and $A\mapsto \bar A$. The last indeed happens for AdS$_3$ if we at the same time map $z\mapsto\bar z$. This means that on the $C$-fields we get the following transformation
\begin{align}\label{eq:autoonC}
 C^s_{m,\sigma}\mapsto(-1)^{-m-s+1}\tilde C^s_{-m,\sigma} \ .
\end{align}

For the Lie superalgebra we define coefficients
\begin{align}
 {g^{(\textrm{Lie})}}^{st}_u (m,n;\lambda,k)&=\frac12 g^{st}_u (m,n;\lambda,k)-(-1)^{4st}\frac12 g^{ts}_u (n,m;\lambda,k) \ .
\end{align}
Using \eqref{eq:conjcoeffi} we get for the bosonic subalgebra
\begin{align}
 {g^{(\textrm{Lie})}}^{st}_u (m,n;\lambda,k)=(-1)^u {g^{(\textrm{Lie})}}^{st}_u (m,n;\lambda,k)\ ,\qquad s,t\in\mathbb{Z}\ ,
\end{align}
and for the anti-commutator of two fermionic operators
\begin{align}
 {g^{(\textrm{Lie})}}^{st}_u (m,n;\lambda,k)=(-1)^{u+1} {g^{(\textrm{Lie})}}^{st}_u (m,n;\lambda,k)\ ,\qquad \textrm{for }s,t\in\mathbb{Z}+1/2\ .
\end{align}
To get a nice result for commutators of bosonic with fermionic operators, we would need to show that the structure coefficients with odd $u$ are independent of $k$, but we will refrain from doing that here.

\subsection{Supertrace}\label{ref:appsupertrace}

In this subsection we will construct the supertrace on the shs$[\lambda]\oplus \mathbb{C}$ Lie superalgebra and show that up to a normalization and one relation it is uniquely determined by the $\mathcal{N}=2$ superalgebra and multiplication with $k$. Put differently, we need to use the invariance under all the generators with spins $1,3/2,2$ and their commutation relations, which were found previously. We will also see that the supertrace has a simple form in terms of the star product. This form will in turn gives us the structure constants $g^{ss}_{2s-1}(m,-m;\lambda, k)$.

An inner product, $\str( ~ , ~ )$, on a Lie superalgebra $\mathcal{G}=\mathcal{G}_0+\mathcal{G}_1$ is defined by \cite{Frappat:1996pb}
\begin{align}
 \str(X,Y)&=0\quad\textrm{for all } X\in\mathcal{G}_0, Y\in\mathcal{G}_1\qquad \textrm{(Consistent)} \nonumber\\
 \str(X,Y)&=(-1)^{\text{deg} X\cdot\text{deg} Y}\str(Y,X)\quad\textrm{for all } X,Y\in\mathcal{G}\qquad \textrm{(Supersymmetric)}\nonumber \\
 \str([X,Y\},Z)&=\str(X,[Y,Z\})\quad\textrm{for all } X,Y,Z\in\mathcal{G}\qquad \textrm{(Invariant)}
\end{align}
where $[ ~ , ~ \}$ denotes the (anti-)commutator. We can now use these properties to explicitly construct the supertrace up to two undetermined constants. The whole subalgebra splits up into supermultiplets which are related by multiplication with $k$. Thus, the invariance of $\mathcal{N}=2$ superalgebra and simple multiplication with $k$ are all that we need besides consistency and (super)symmetry to determine the supertrace.

Basically, we want to determine $\str\left(V^{s\sigma}_m,V^{s'\sigma'}_{m'}\right)$. The invariance under the sl(2) subalgebra $V^{2+}_m$, with commutation relations given in \eqref{eq:commuosp12}, gives us
\begin{align}
\str\left(V^{s\sigma}_m,V^{s'\sigma'}_{m'}\right)&\propto\delta_{m,-m'}\delta_{s,s'}\ ,\nonumber \\
\str\left(V^{s\sigma}_m,V^{s\sigma'}_{-m}\right)&=(-1)^m\frac{(m+s-1)!(s-m-1)!}{((s-1)!)^2}\str\left(V^{s\sigma}_0,V^{s\sigma'}_{0}\right)\quad\textrm{for }s\in\mathbb{Z}\ ,\label{eq:mdependencestr}\\
 \str\left(V^{s\sigma}_m,V^{s\sigma'}_{-m}\right)&=(-1)^{m-1/2}\frac{(m+s-1)!(s-m-1)!}{((s-3/2)!)^2(s-1/2)}\str\left(V^{s\sigma}_{1/2},V^{s\sigma'}_{-1/2}\right)\quad\textrm{for }s\in\mathbb{Z}+1/2\ .\nonumber
\end{align}
Next we need to know what the dependence on $k$ is. First for the fermionic part we use $[k,V^s_m]=2kV^s_m$ and invariance to get
\begin{align}
 \str\left(kV^{s}_m,kV^{s}_{-m}\right)&=-\str\left(V^{s}_m,V^{s}_{-m}\right)\quad\textrm{for }s\in\mathbb{Z}+1/2\ ,\\
\str\left(k V^{s}_m,V^{s}_{-m}\right)&=0\quad\textrm{for }s\in\mathbb{Z}+1/2\ ,
\end{align}
where in the last equation we have used that the supertrace is anti-symmetric in the fermionic generators, and that it is an odd function in the $m$-labels for fermionic generators, see eq. \eqref{eq:mdependencestr}. For the bosonic part we need the multiplication with $k$. That is, we also use the following commutators $kV^s_m=-\frac1m [k V^2_0,V^s_m]$ and $[kV^2_0,kV^s_m]=[V^2_0,V^s_m]$ together with invariance under $kV^2_0$. Hence, we are using the invariance of all generators of spin $1,3/2,2$. We then get
\begin{align}
\str\left(k V^{s}_m,kV^{s}_{-m}\right)&=\str\left( V^{s}_m,V^{s}_{-m}\right)\quad\textrm{for }s\in\mathbb{Z}\textrm{ and }s>1\ .
\end{align}
Further, we use that $\{kV^{s+1/2}_{m-1/2},V^{3/2}_{1/2}\}=k[V^{s+1/2}_{m-1/2},V^{3/2}_{1/2}]$ and using the explicit star products calculated above, we get
\begin{align}
 \str\left(k V^{s}_m,V^{s}_{-m}\right)&=\frac{-\nu}{2s-1}\str\left( V^{s}_m,V^{s}_{-m}\right)\quad\textrm{for }s\in\mathbb{Z}\ .
\end{align}
Since we have a non-trivial ideal being the span of the identity operator, we have to determine the normalization of $\str(1,1)$ together with the normalization of say $\str(k,k)$. We will make a star product construction of the supertrace, so with this in mind the most natural choice is $\str(k,k)=\str(1,1)$ leaving only the overall normalization undetermined.

The supertrace is finally determined using the stepping relation coming from invariance under $V^{3/2}_m$. Using the above result we get
\begin{align}
  \str\left( V^{s}_m,V^{s}_{-m}\right)&=\frac14(m+(s-1))\str\left( V^{s-1/2}_{m-1/2},V^{s-1/2}_{-m+1/2}\right)\quad\textrm{for }s\in\mathbb{Z}\ ,\\
  \str\left( V^{s}_m,V^{s}_{-m}\right)&=\frac14(m+(s-1))(1-\frac{\nu^2}{4(s-1)^2})\str\left( V^{s-1/2}_{m-1/2},V^{s-1/2}_{-m+1/2}\right)\quad\textrm{for }s\in\mathbb{Z}+1/2 \nonumber
\end{align}
with the solution (presented in the form using the projection onto $k$-eigenspaces)
\begin{align}\label{eq:supertraceexpliciteven}
 &\str\left(P^\pm V^{s}_m,P^\pm V^{s}_{-m}\right)  \\
& =\frac{(-1)^{s-m-1}\Gamma(s+m)\Gamma(s-m)}{(2s-2)!}\frac{\Gamma(s)\sqrt{\pi}}{4^s\Gamma(s+1/2)}(1-\lambda_\pm)_{s-1}(1+\lambda_\pm)_{s-1}\lambda_\pm \nonumber
\end{align}
for $s \in \mathbb{Z}$ and
\begin{align}\label{eq:supertraceexplicitodd}
 &\str\left(P^\pm V^{s}_m,P^\mp V^{s}_{-m}\right) \\
&=\frac{(-1)^{s-m-1}\Gamma(s+m)\Gamma(s-m)}{(2s-2)!}\frac{\Gamma(s-\tfrac12)\sqrt{\pi}}{4^s\Gamma(s)}(1-\lambda_+)_{s-\frac12}(1+\lambda_+)_{s-\frac32}\lambda_+\nonumber
\end{align}
for $s \in \mathbb{Z}+1/2$,
where we used the ascending Pochhammer symbol $(a)_n=\Gamma(a+n)/\Gamma(a)$, and for simplicity have taken the normalization $\str(1,1)=1$. This indeed has a form similar to the invariant metric suggested in \cite{FL}, and the bosonic case gives the same result as in \cite{Gaberdiel:2011wb} eq. (A.3) with $q=1/4$ (and remembering the different overall normalization).

We can now show that such an inner product indeed exists and has the following star product form
\begin{align}\label{eq:appsupertracestarproduct}
 \str\big(V^{s\sigma}_m,V^{s'\sigma'}_{m'}\big)=2\lambda_k*V^{s\sigma}_m*V^{s'\sigma'}_{m'}\big|_1\ ,
\end{align}
where the projection is onto the span of the identity operator. Here $\lambda_k=(1-\nu k)/2$ as before, and we have normalized such that $\str(1,1)=1$. We are here of course forced to have $\str(k,k)=\str(1,1)$. This is immediately consistent, and we also see that the spins of the two operators have to be the same, and the $m$-numbers have to be opposite. If we can show supersymmetry, invariance will follow immediately via the definition of the star-supercommutator. Supersymmetry is almost determined by the automorphism $\sigma$ \eqref{eq:anti-auto}:
\begin{align}
 2\lambda_k*V^{s\delta}_m*V^{s'\delta'}_{m'}\big|_1=\sigma(2\lambda_k*V^{s\delta}_m*V^{s'\delta'}_{m'})\big|_1=(-1)^{s+s'-2}(\delta)^{2s+1}(\delta')^{2s'+1}2\lambda_k*V^{s'\delta'}_{m'}*V^{s\delta}_m\big|_1 \ . \nonumber
\end{align}
This shows symmetry in the bosonic case and anti-symmetry in the fermionic case when $\delta=\delta'$. Since it says that we have symmetry in the fermionic case when $\delta\neq\delta'$, we need to show that we here get zero.
As we have also seen above, it will be a consequence of the anti-symmetry. To show this we first see explicitly that it is true for the spin $3/2$ part; $2\lambda_k*V^{3/2}_m*kV^{3/2}_{-m}\big|_1=0$. This means that we have invariance for the supercharges. This gives us the wanted result (assuming here for simplicity $m\neq s-1/2$)
\begin{align}
 \str\big(kV^{s}_m,V^{s}_{-m}\big)&=\frac{-2(s-1)}{(m-s+1/2)(s-1/2)}\str\big(kV^{s}_m,[V^{3/2}_{1/2},V^{s+1/2}_{-m-1/2}]\big)\nonumber\\
 &\propto \str\big(\{kV^{s}_m,V^{3/2}_{1/2}\},V^{s+1/2}_{-m-1/2}\big)=\str\big(k[V^{s}_m,V^{3/2}_{1/2}],V^{s+1/2}_{-m-1/2}\big)=0 \ .
\end{align}
We thus have supersymmetry and the explicit equations for the supertrace above applies.

On the other hand the star product formula for the supertrace means that
\begin{align}\label{eq:strstructure}
    \str\big(P_\sigma V^{s+}_m,P_{\sigma'}V^{s'+}_{m'}\big)&=\frac12\delta_{\sigma,(-1)^{2s}\sigma'}\delta_{s,s'}\delta_{m,-m'}\lambda_\sigma g^{ss}_{2s-1}(m,-m;\lambda,k=\sigma1)\ ,
\end{align}
which gives us explicit formulas for the structure constants
\begin{align}
& g^{ss}_{2s-1}(m,-m;\lambda,k=\sigma1) \\
&=\frac{(-1)^{s-m-1}\Gamma(s+m)\Gamma(s-m)}{(2s-2)!}\frac{2\Gamma(s)\sqrt{\pi}}{4^s\Gamma(s+1/2)}(1-\lambda_\sigma)_{s-1}(1+\lambda_\sigma)_{s-1}\nonumber
\end{align}
for $s\in\mathbb{Z}$ and
\begin{align}
& g^{ss}_{2s-1}(m,-m;\lambda,k=\sigma1) \\
&=\frac{(-1)^{s-m-1}\Gamma(s+m)\Gamma(s-m)}{(2s-2)!}\frac{2\Gamma(s-\tfrac12)\sqrt{\pi}}{4^s\Gamma(s)}(1-\lambda_\sigma)_{s-\frac12}(1+\lambda_\sigma)_{s-\frac32}\nonumber
\end{align}
for $s\in\mathbb{Z}+1/2$.

\subsection{Bulk field couplings}

When we want to calculate two-point functions, we need to know how the fields couple. For this we consider the simplest possible non-trivial action which is gauge invariant under \eqref{eq:gaugetransoriginalfields}, which is the
mass-like term
\begin{align}\label{}
    S=A\int d^3 x \sqrt{G}\int d\psi_1\psi_1\int d\psi_2\psi_2\str \big(\mathcal{C}*\mathcal{C}\big)+\textrm{c.c.}\ .
\end{align}
In the bosonic case the trace is defined as the restriction of the star product to the constant part, however in the supersymmetric case we have to be a bit more careful. As shown in the previous subsection, we define the supertrace as (see eq. \eqref{eq:appsupertracestarproduct})
\begin{align}
 \str\big(V^{s\sigma}_m,V^{s'\sigma'}_{m'}\big)=2\lambda_k*V^{s\sigma}_m*V^{s'\sigma'}_{m'}\big|_1\ ,
\end{align}
where $\lambda_k=(1-\nu k)/2$. Since we have an ideal generated by the identity operator, we have to fix two normalizations in the supertrace, in particular we have here chosen $\str(k,k)=\str(1,1)=1$. An explicit formula for the supertrace can be found using the invariance under the generators with spin 1, 3/2 and 2, see eqs. \eqref{eq:supertraceexpliciteven}, \eqref{eq:supertraceexplicitodd}. To keep things short we here simply write as in \eqref{eq:strstructure}.
We can then write the action out into components as
\begin{align}\label{eq:lagrangianstructure}
\begin{split}
    \mathcal{L}=&\frac{A}{2}\sum_{s=1,2,\ldots}\sum_{ |m|\leq s-1}\sum_{\sigma=\pm} C^s_{-m, \sigma}\tilde C^s_{m, \sigma}\lambda_\sigma g^{ss}_{2s-1}(m,-m;\lambda,k=\sigma1)\\
    +& \frac{A}{2}\sum_{s=3/2,5/2,\ldots}\sum_{ |m|\leq s-1}\sum_{\sigma=\pm} C^s_{m, \sigma}\tilde C^s_{-m, -\sigma}\lambda_\sigma g^{ss}_{2s-1}(m,-m;\lambda,k=\sigma1)+\textrm{c.c.} \ .
\end{split}
\end{align}
This is indeed invariant under the anti-automorphism $\eta$ defined in \eqref{eq:anti-auto} which sends $\eta(C^s_{m,\sigma})=(-1)^{-s+1}\tilde C^s_{m,(-1)^{2s}\sigma}$ using the symmetries of the structure constants. It is also invariant under the automorphism taking $C^s_{m,\sigma}\mapsto(-1)^{m+s-1}\tilde C^s_{-m,\sigma}$.

\section{CFT OPEs and commutator relations}
\label{CFTW}

\subsection{$\mathcal{N}=2$ superconformal algebra}

The $\mathcal N=2$ chiral superconformal algebra with Virasoro central charge $c$ has the form
\begin{align}\label{eq:superconfalgope}
    G^+(z)G^-(w)&\sim\frac{2c/3}{(z-w)^3}+\frac{2U(w)}{(z-w)^2}+\frac{2T(w)+\del U(w)}{z-w} \ ,\nonumber \\
    G^\pm(z)G^\pm(w)&\sim 0 \ ,\nonumber \\
    T(z)T(w)&\sim\frac{c/2}{(z-w)^4}+\frac{2T(w)}{(z-w)^2}+\frac{\del T(w)}{z-w} \ ,\nonumber \\
    T(z)G^\pm(w)&\sim\frac{\tfrac{3}{2}G^\pm(w)}{(z-w)^2}+\frac{\del G^\pm(w)}{z-w} \ ,\nonumber \\
    T(z)U(w)&\sim\frac{U(w)}{(z-w)^2}+\frac{\del U(w)}{z-w} \ ,\nonumber \\
    U(z)U(w)&\sim\frac{c/3}{(z-w)^2} \ ,\nonumber \\
    U(z)G^\pm(w)&\sim\pm\frac{G^\pm(w)}{z-w}
\end{align}
or in terms of generators
\begin{align}\label{eq:superconfalg}
\begin{split}
[L_m,L_n]&=(m-n)L_{m+n}+\frac{c}{12}(m^3-m)\delta_{m,-n}\ ,\\
[L_m,G_r^\pm]&=(m/2-r)G^\pm_{m+r}\ , \\
[L_m,U_n]&=-nU_{m+n}\ , \\
\{G^+_r,G^-_s\}&=2L_{r+s}+(r-s)U_{r+s}+\frac{c}{3}(r^2-\frac14)\delta_{r,-s}\ ,\\
\{G^\pm_r,G^\pm_s\}&=0\ ,\\
[U_m,G^\pm_r]&=\pm G^\pm_{m+r}\ ,\\
[U_m,U_n]&=\frac{c}{3}m\delta_{m,-n}\ .
\end{split}
\end{align}

\subsection{${\cal W}$ algebra}

Assuming that we have an $\mathcal N=2$ supersymmetric ${\cal W}$ algebra, we have supermultiplets $(W^{s0},W^{s\pm},W^{s,1})$ where (see e.g. \cite{Candu:2012tr})
\begin{align}
 W^{s\pm}=\mp G^{\pm}_{-1/2}W^{s0}\ ,\qquad W^{s1}=\frac{1}{4}(G^{+}_{-1/2}G^{-}_{-1/2}-G^{-}_{-1/2}G^{+}_{-1/2})W^{s0}\ .
\end{align}
The combination in the last equation ensures that we have chiral primaries, and have been chosen such that $W^{(1)0}=U$, $W^{(1)\pm}=G^{\pm}$ and $W^{(1)1}=T$. For each bosonic spin (except spin one) we thus have two higher spin fields $W^{s0}$ and $W^{(s-1)1}$, where the field $W^{s0}$ has U(1)-charge zero.  The corresponding OPEs are then
\begin{align}\label{eq:susytranswalg}
\begin{split}
  G^\pm(z)W^{s0}(w)&\sim\mp\frac{W^{s\pm}(w)}{z-w}\ ,\\
  G^\pm(z)W^{s\pm}(w)&\sim0\ ,\\
  G^\pm(z)W^{s\mp}(w)&\sim\pm\frac{2s W^{s0}(w)}{(z-w)^2}+\frac{ 2 W^{s1}(w)\pm\del W^{s0}}{z-w}\ ,\\
  G^\pm(z)W^{s1}(w)&\sim\frac12\frac{(2s+1) W^{s\pm}(w)}{(z-w)^2}+\frac12\frac{\del W^{s\pm}}{z-w}\ ,\\
  U(z)W^{s0}(w)&\sim0\ ,\\
  U(z)W^{s1}(w)&\sim \frac12 s\frac{W^{s0}(w)}{(z-w)^2}\ .
\end{split}
\end{align}

\subsection{States}\label{sec:states}

A chiral state
\begin{align}
 G^+(z)\phi_h(w)\sim 0
\end{align}
fulfills $2L_0=U_0$ and its superpartner $\psi_h$
\begin{align}
 G^-(z)\phi_h(w)\sim\frac{\psi_h(w)}{z-w}
\end{align}
has OPEs
\begin{align}\label{eq:gpluschiralope}
 G^+(z)\psi_h(w)\sim \frac{4h\phi_h}{(z-w)^2}+\frac{2\del\phi_h}{z-w}\ ,\quad G^-(z)\psi_h(w)\sim0\ .
\end{align}
An anti-chiral state
\begin{align}
 G^-(z)\tilde\phi_h(w)\sim 0
\end{align}
similarly fulfills $2L_0=-U_0$ and its superpartner $\tilde\psi_h$
\begin{align}
 G^+(z)\tilde\phi_h(w)\sim\frac{\tilde\psi_h(w)}{z-w}
\end{align}
has OPEs
\begin{align}
 G^-(z)\tilde\psi_h(w)\sim \frac{4h\tilde\phi_h}{(z-w)^2}+\frac{2\del\tilde\phi_h}{z-w}\ ,\quad G^+(z)\tilde\psi_h(w)\sim0\ .
\end{align}

\end{document}